\pdfoutput=1
\documentclass[useAMS,usenatbib]{mn2e}

\usepackage{amssymb}
\usepackage{amsmath}
\usepackage[dvipdfmx]{graphicx}	
\usepackage{bmpsize}
\usepackage{float}
\usepackage{tabulary}
\usepackage{rotating}
\usepackage{caption}
\usepackage{cmap}
\usepackage[usenames, dvipsnames]{color}
\usepackage{lscape}
\usepackage{pdflscape}
\usepackage[titletoc,title]{appendix}
\usepackage{soul}

\bibpunct{(}{)}{;}{a}{,}{,}

\newcommand{\corrnobf}[1]{{#1}}

\newcommand{\corr}[1]{{#1}}

\newcommand{\corrdre}[1]{{#1}}

\definecolor{darkgreen}{rgb}{0.1, 0.6, 0.1}
\title[3D Hydrodynamic Simulations of Carbon Burning in Massive Stars]{3D Hydrodynamic Simulations of Carbon 
Burning in Massive Stars}
\author[A. Cristini, C. Meakin, R. Hirschi, D. Arnett, C. Georgy, M. Viallet, I. Walkington]{A. Cristini$^{a}$\thanks{E-mail: a.j.cristini@keele.ac.uk}, C. Meakin$^{b,c}$, R. Hirschi$^{a,d}$, D. Arnett$^c$, 
C. Georgy$^{e,a}$, M. Viallet$^{f}$,\\\\ \hspace{-1mm}\LARGE{\textnormal{I. Walkington$^{a}$}}\\\\
$^a$\textsf{Astrophysics Group, Keele University, Lennard-Jones Laboratories, Keele, ST5 5BG, UK}\\
$^b$\textsf{Karagozian \& Case, Inc., 700 N. Brand Blvd. Suite 700, Glendale, CA, 91203, USA}\\
$^c$\textsf{Department of Astronomy, University of Arizona, Tucson, AZ 85721, USA}\\
$^d$\textsf{Kavli IPMU (WPI), The University of Tokyo, Kashiwa, Chiba 277-8583, Japan}\\
$^e$\textsf{Geneva Observatory, University of Geneva, Ch. Maillettes 51, 1290 Versoix, Switzerland}\\
$^f$\textsf{Max-Planck-Institut f\"{u}r Astrophysik, Karl Schwarzschild Strasse 1, Garching, D-85741, Germany}}
\vspace{-1cm}
\begin{document}

\date{Accepted 15/06/2017. Received 17/10/2016}

\pagerange{\pageref{firstpage}--\pageref{lastpage}} \pubyear{2017}

\maketitle

\label{firstpage}

\begin{abstract} 
We present the first detailed three-dimensional (3D) hydrodynamic implicit large eddy simulations 
of turbulent convection of carbon burning in massive stars. Simulations begin with
radial profiles mapped from a carbon burning shell within a 15$\,\textrm{M}_\odot$ one-dimensional stellar evolution model.
\corr{We consider models with $128^3$, $256^3$, $512^3$ and $1024^3$ zones. }
The turbulent flow properties of these 
carbon burning simulations are very similar to the oxygen burning case. 
We performed a mean field analysis of the kinetic 
energy 
budgets within the Reynolds-averaged Navier-Stokes framework.
For the 
upper convective boundary region, we find that the
numerical dissipation is insensitive to 
resolution for linear mesh resolutions above 512 grid points.
For the stiffer, more stratified 
lower boundary, our highest resolution model still shows signs of decreasing \corr{sub-grid} dissipation 
suggesting it is not yet numerically converged.
We find that the widths of the upper and lower boundaries are roughly 30\% and 10\% of the local 
pressure scale heights, respectively. 
The shape of the boundaries is significantly different from those used in
stellar evolution models.
\corr{ As in past oxygen-shell burning simulations, we observe entrainment at both boundaries in our carbon-shell burning simulations. 
In the large P\'eclet number regime found in the advanced phases, the entrainment rate is roughly inversely proportional to the 
bulk Richardson number, Ri$_{\rm B}$ ($\propto $Ri${\rm_B}^{-\alpha}$, $0.5\lesssim \alpha \lesssim 1.0$). 
We thus suggest the use of Ri$_{\rm B}$ as a means to take into account the results of 3D hydrodynamics
simulations in new 1D prescriptions of convective boundary mixing.}\\\
\end{abstract}


\section{Introduction}\label{intro}



\par One-dimensional (1D) stellar evolution codes are currently the only way to simulate the entire lifespan of a star. 
This comes at the cost of having to replace complex, inherently three-dimensional (3D) processes, such as convection, 
rotation and magnetic activity, with generally simplified mean-field models.
An essential question is {\em ``how well do these 1D models represent reality?''} Answers can be found both in empirical
and theoretical work.  On the empirical front, we can investigate full star models, by comparing them to observations of stars
under a range of conditions, as well 
as testing the basic physics that goes into models of multi-dimensional phenomena
by studying relevant laboratory work \corr{ and data from meteorology and oceanography (remembering that stars are much bigger than planets, and are composed of high energy-density plasma). }
On the theoretical side, multi-dimensional simulations can be used to test 
1D models under astrophysical conditions that \corr{ can be recreated in terrestrial laboratories only in small volumes e.g. in NIF \citep{2011ApSS..336..207K} and z-pinch device \citep{2013AcAau..82..173M} experiments.}

\subsection{Astronomical tests}
\par The results from the \corr{ astronomical} validation studies are mixed.  Observations of stars confirm the general, {\em qualitative} picture 
of stellar evolution predicted by 1D models, but reveal significant {\em quantitative} differences.  A recent example is the work
of  \citet{2014MNRAS.439L...6G} and \citet{2013A&A...560A..16M} who show that the use of different criteria for convection 
(i.e., either Schwarzchild or Ledoux) leads to important differences in the overall evolution of a massive star, 
especially for the post-main-sequence evolution. Without a constraint on which
criteria, if either, is the correct one, this result represents an inherent uncertainty in 1D models. 

\par These quantitative discrepancies can be reduced by modifying the treatment of convective boundaries, and more specifically, by 
allowing for convective penetration and overshooting \citep{1991A&A...252..179Z}.  Incorporating a model for mixing beyond the linearly stable convective 
boundaries (e.\,g. that given by the Ledoux or Schwarzchild criteria) introduces additional parameters that can be tuned to 
improve agreement between model and data \citep{1996A&A...313..497F}.  However, this approach has several drawbacks beyond the
obvious one of over-fitting so as to preclude a predictive model.  Perhaps the most egregious is that parameter fitting is never done
in a global sense so that different phases of evolution require different parameters, thus revealing the non-universality of these models. \corr{Another recent example is the finding that stellar models of red giants agree with Kepler observations only when a metallicity dependant mixing length is used \citep{2017arXiv170401164T}.}

\subsection{Computational Methods and Assumptions}

\par \corr{The most obvious way to  proceed computationally is by direct numerical simulation (DNS), in which all relevant scales of the turbulent cascade are resolved. This is not feasible with present or foreseeable computer power. The Reynolds numbers for stars are enormous \citep[e.g. ${\rm Re}\approx 10^{18}$][]{2014AIPA....4d1010A}, simply because stellar dimensions are so much larger than mean-free-paths for dissipation. DNS requires an infeasible dynamic range in order to include both the microscopic and macroscopic scales; for example, the state-of-the-art DNS work of \cite{2013JFM...732..150J}
attained a Reynolds number of $10^3$ with a P\'eclet number of unity.}

\corr{An alternative is possible. The largest eddies contain most of the  energy  in a turbulent cascade. Kolmogorov's second similarity hypothesis, which posits that the rate of dissipation in a turbulent flow as well as the statistics in the inertial sub-range do not depend upon the detailed nature of the dissipative process, implies that it may be unnecessary to resolve the dissipation sub-range to accurately calculate scales above the Kolmogorov scale, provided that the behaviour of the sub-grid dissipation is well behaved. This phenomenology has indeed been supported by detailed numerical studies \citep[see][]{Aspden2008}. Even early ILES simulations with relatively coarse resolution \citep{2007ApJ...667..448M}
 gave Kolmogorov dissipation at the sub-grid scale; this is because they use a finite volume and total variations diminishing (TVD) solver \citep[PPM; see][]{1984JCoPh..54..174C}, ensuring that mass, momentum, and energy are conserved and variance is dissipated at the grid scale.}
 
\par \corr{Comparative studies of using DNS to solve the compressible Navier-Stokes equations and ILES to solve the inviscid Euler equations using PPM have been performed \citep{2000ApJS..127..159P,2000JCoPh.158..225S}. Comparisons were made on grids with sizes from $64^3$ to $1024^3$. Both methods were found to converge to the same limit with increasing resolution.} \corr{A factor in deciding whether DNS or ILES is a more suitable choice depends on whether the phenomena of interest require resolution of the dissipative range or not. We currently do not have a compelling argument for resolving the dissipation range in the current work.}

\par \corr{Furthermore, the additional information provided explicitly by DNS, such as dissipation rates, can often be estimated very accurately when the ILES method is used in conjunction with Reynolds-averaged Navier Stokes (RANS) methods; at least in the mean. This is a point we discuss in \S~\ref{MFA} below and in \cite{2013ApJ...769....1V,2015ApJ...809...30A} and \citet{2016RPPh...79j2901A}.}

\subsection{Stellar simulations}

\par ILES simulations sampling a broad range of relevant and increasingly more realistic 
astrophysics conditions have been undertaken.
\corr{ Neutrino cooling becomes dominant after helium burning, so that later stages have increasingly shorter thermal time-scales \citep[see pg. 284 - 292 of][]{1996snih.book.....A}, which are insensitive to radiative diffusion or heat conduction (high P\'eclet number}\footnote{The P\'{e}clet number is the ratio of the time-scale for transport of heat through conduction to the time-scale for transport of heat through advection, or $\textrm{Pe}=vL/\chi$, 
where $v$ and $L$ are the characteristic velocity and length-scale of the flow and $\chi$ is the heat diffusivity 
\citep[e.\,g. pg. 380 of][]{lautrup_2011}.}, \corr{${\rm Pe} \gg 1$). Oxygen burning has both a relatively simple nuclear burning process, and a short thermal time, so that a small but significant fraction of the burning stage may be simulated \citep{2007ApJ...667..448M},
with a Damk\"{o}hler number}\footnote{The Damk\"{o}hler number is the
ratio of the advective time-scale to the chemical/nuclear time-scale \citep{1940zfe...46.11.601}, or $\textrm{Da}=\tau_c\,/(qX_i/\epsilon_{nuc})$, where $\tau_c$ is the convective turnover time and $\epsilon_{nuc}$, $q$ and $X_i$ are the energy generation rate, specific energy released and abundance fraction for the dominant nuclear reaction, respectively.},  \corr{Da, approaching 1\% (see 
Table\,\ref{param} for estimates of Da for various burning stages).}

\corr{Many oxygen burning simulations have 
been performed, giving an improved understanding of the process; e.g.,
\citet{1994ApJ...427..932A,1994ApJ...433L..41B,1998ApJ...496..316B,2000ApJ...545..435A,2003ASPC..293..147K,2005ApJ...629L.101Y,2006ApJ...637L..53M,2007ApJ...665..690M,2007ApJ...667..448M,2011ApJ...733...78A,2013ApJ...769....1V,2015ApJ...809...30A,2016RPPh...79j2901A,2017MNRAS.465.2991J}.}

\par \corr{Silicon burning is the most complex burning phase, complicated by active nuclear weak interactions,  
and requires a large additional computational effort. 
The evolution time-scale is of
 the order of days ($\rm Da \sim 1, \rm Pe \gg 1$). Early simulations of silicon burning  \citep{1997NuPhA.621..607B} used a nuclear reaction network consisting of 123 nuclei.
\corr{\citet{2006PhDT........CM} and \citet{2011ApJ...733...78A} performed 2D simulations of concentric carbon, oxygen, and silicon burning shells using a 37 species network for several convective turnovers about one hour prior to core collapse.}
\citet{2015ApJ...808L..21C} simulate the final three minutes of silicon burning 
in a 15$\,\textrm{M}_\odot$  star, using the \textsc{flash} 
code \citep{2000ApJS..131..273F} with adaptive mesh refinement, and a nuclear reaction network of 21 species.
An initial  study of silicon burning with a large network ($\sim 120$ nuclei) has been carried out by 
Meakin \& Arnett (in prep.). The carbon, 
oxygen and part of the silicon shell of an 18$\,\rm M_\odot$, unrelaxed spherical star have also been simulated, 
in a full-sphere simulation with low resolution ($400\times148\times56$) 
by \citet[]{2016ApJ...833..124M}.}

\par \corr{Early phases of stellar evolution are harder to simulate because they are generally characterised by very small 
Damk\"{o}hler numbers (slow burning) and very low convective Mach numbers (slow mixing).  Several studies have targeted hydrogen or helium burning phases.
\citet{2007ApJ...667..448M} performed a fully-compressible simulation of core 
hydrogen burning on a numerical grid of $400\times100^2$, with the driving luminosity boosted by a factor of 10. 
\citet{2013ApJ...773..137G} adopt the low Mach number solver \textsc{Maestro} \citep{2007JPhCS..78a2085A} to simulate 
core hydrogen burning on a numerical grid of $512^3$. This type of solver removes the need to follow the propagation of 
acoustic waves, and allows for longer time-steps than a fully compressible solver, but would neglect any important 
kinetic energy transfer due to acoustic fluxes.}

\par \corr{We have performed novel calculations of a yet to be simulated phase of evolution, the carbon 
phase in a massive star, which we studied in a burning shell within a 15 $\rm M_\odot$ massive star.  
Carbon burning is the first neutrino-cooled  burning 
stage, thus allowing radiative diffusion to be neglected ($\rm Pe \gg 1$) and slightly simplifying
the numerical model. It is characterised by a larger  Damk\"{o}hler number than earlier, radiatively cooled stages, 
alleviating the computational cost. The initial composition and structure profiles are simpler than those of more advanced stages, because 
the region in which the shell forms is smoothed by the preceding convective helium-burning core. 
Finally, as the first neutrino dominated 
phase of nuclear burning it plays an important role is setting the size of the heavy element core which subsequently forms and in which a 
potential core-collapse event may take place.
We are particularly interested in  the structure of convective boundaries and composition gradients, in this sense we explore the effects of resolution (zoning) upon the simulations. Composition is treated as an active scalar, and coupled to the fluid flow through advection and the equation of state (EOS).}

\par The structure of the paper is as follows. In \S\ref{phys} we discuss the stellar model from which the initial conditions
for our hydrodynamic models were selected. In \S\ref{method} we describe our simulation model set-up. Our results and 
analysis of the hydrodynamic models are presented in \S\ref{results}. We compare our models
to similar simulations in \S\ref{other_sims}. Finally, in \S\ref{conc} we summarise our results.

\section{Initial Conditions}\label{phys}

\subsection{The 1D Stellar Evolution Model}\label{model}

\par To prepare the input for the 3D carbon burning simulations, we calculated a 15$\,\textrm{M}_\odot$, solar metallicity, non-rotating model 
until the end of the oxygen burning phase using the Geneva stellar evolution  code \citep[\textsc{genec};][]{2008Ap&SS.316...43E}.  The 
default input physics used in \textsc{genec} to calculate this model includes: a nuclear reaction network of 23 isotopes using the NACRE \citep{1999NuPhA.656....3A} tabulated reaction rates; EOS describing a perfect gas, partial degeneracy and radiation; opacity 
tables from the OPAL group \citep{1996ApJ...456..902R} and \citet{1994ApJ...437..879A} for high and low temperatures, respectively; 
mass loss estimated according to the prescriptions by \citet{2001A&A...369..574V} and \citet{1988A&AS...72..259D}; concentration 
and thermal diffusion; convection treatment using MLT with $\alpha_{ml} = 1.6$ \citep{1992A&AS...96..269S}; convective boundary positions 
determined using the Schwarzschild criterion \citep{1992gwcw.bookR....S}; and penetrative convective overshoot \citep{1991A&A...252..179Z} up 
to 20$\%$ \citep{1991ApJ...381L..67S} of the pressure scale height for core hydrogen and helium burning only.

\begin{figure}
\hspace{-7mm}\includegraphics[width=0.55\textwidth,height=0.35\textwidth]{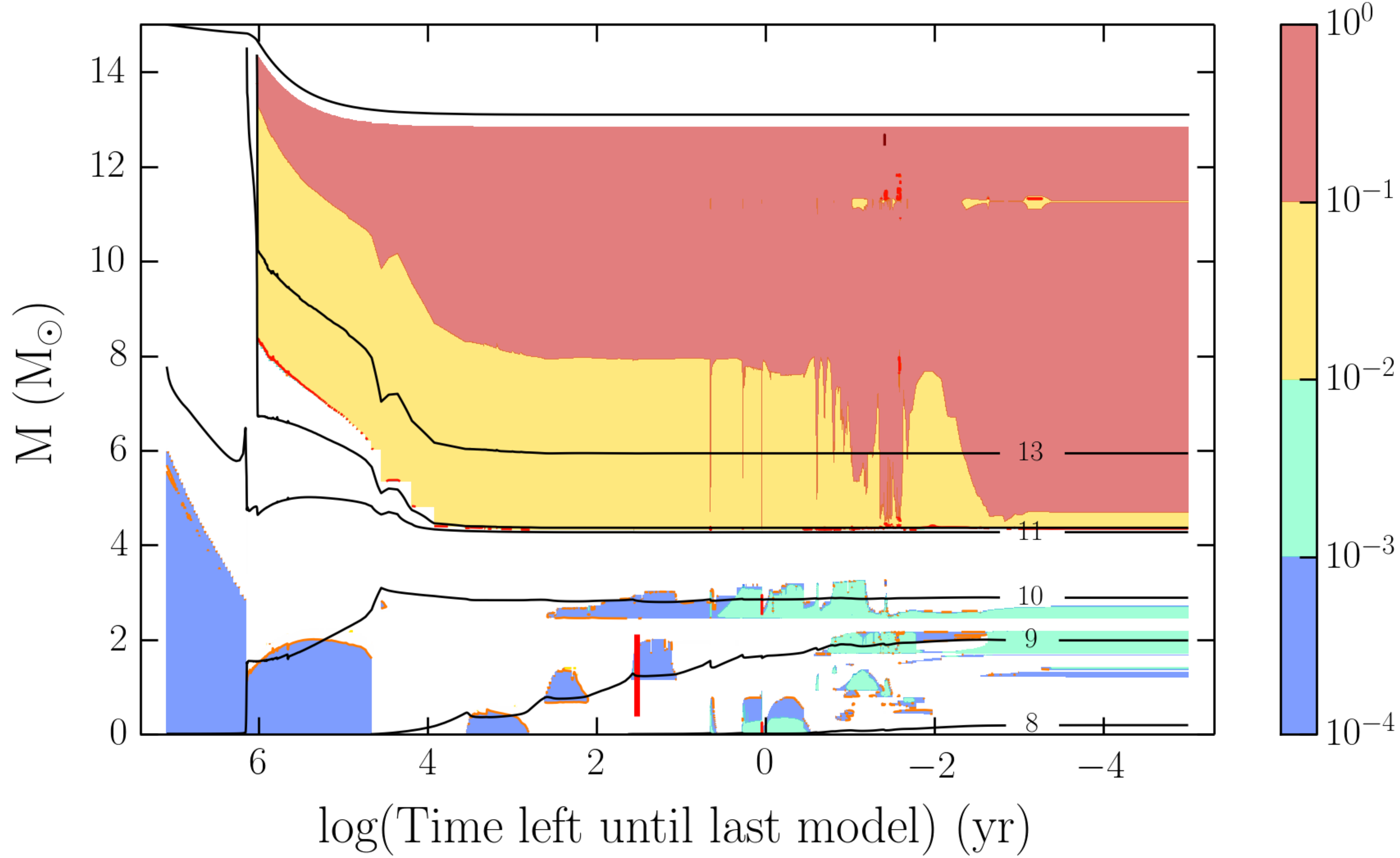}
\caption{Structure evolution diagram of the 15$\,\textrm{M}_\odot$ 1D input stellar model. The horizontal axis is a 
logarithmic scale of the time left 
before the predicted collapse of the star in years (the last model in this simulation is before the end of silicon 
burning but since the time-scale of silicon burning is so short this does not affect the plot for the earlier phases) 
and the vertical axis is the mass in solar masses. The total mass and radial contours (in the form 
$\textrm{log}_{10}(\textrm{r})$ in cm), are drawn as solid black lines. Shaded areas correspond to convective regions. The colour indicates 
the value of the Mach number. The red vertical bar around log{\rm [time left in years]}$\sim$ 1.5 represents the domain 
simulated in 3D, and the time at which the 3D simulations start, \corr{relative to the evolution of the star}.}
\label{kipp}
\end{figure} 

\par Figure \ref{kipp} presents the evolution of the convective structure of this 15$\,\rm {M}_{\odot}$ model. 
Convectively unstable regions are indicated in this figure by shaded areas with colour indicating the convective Mach number, 
\corr{which slowly rises as the star evolves, being lowest in the core and highest in the envelope.}

\subsection{An Overview of Stellar Convection Parameters}\label{sc}

\par  In order to place the results of our carbon shell simulations into the broader context of stellar convection over 
the lifetime of the star, as well as inform the \corr{ construction of initial states for}
future simulations, we have estimated key quantities for 
most of the convective zones
in the 15$\,\rm M_{\odot}$ model (Fig. \ref{kipp}). These quantities include the bulk Richardson number, $\textrm{Ri}_{\rm B}$ (Eq. \ref{rib}); 
convective velocity, $v_c$ (Eq. \ref{vc}); Mach number, $\textrm{Ma}$ (Eq. \ref{ma}); P\'{e}clet number, $\textrm{Pe}$ (Eq. \ref{pe}); and Damk\"{o}hler number, $\textrm{Da}$ (Eq. \ref{da}). These values and the methods by which they have been calculated are presented in Appendix 
\ref{convection-parameters-appendix}. 
\corr{These are order of magnitude estimates intended to show trends between different stages of evolution.} 

\par One additional key property of the advanced convective regions in massive stars is the radial extent (see the radial 
contours in Fig. \ref{kipp}). For the mass range that we consider, such convective regions typically span only a 
few pressure scale heights ($0.2-5$), convection, in this case, is classified as shallow\footnote{An example of deep 
convection is in the 
envelopes of red giants, which extends over many pressure scale heights.}. 
Consequently, convective motions might be expected to resemble at least some characteristics of the classical 
description of convective rolls proposed by \citet{1963JAtS...20..130L}, a hypothesis that shows some validity according 
to the results of \citet{2011ApJ...741...33A}.

\par Referring to Table~\ref{param}, \corr{ the 1D model (shown in Fig. \ref{kipp}) shows a general increase in the convective velocities and the Mach, P\'eclet and Damk\"ohler numbers as the star evolves.} Some additional trends of interest include the following. \\

\noindent {\em Convective Velocity. --- }The convective velocities range from about $5\times10^4$\,cm$\,$s$^{-1}$ during 
the early phases to a few times  $10^6$\,cm\,s$^{-1}$ during the advanced phases.\\

\noindent {\em Mach Number. ---} The Mach number ranges from a few times $10^{-4}$ (values lowest for helium and carbon burning) to close to $10^{-2}$ \corr{ (several times $10^{-2}$ for 3D simulations). } Note that the Mach number may 
still increase further during silicon burning and the early collapse as found by \citet{1996snih.book.....A} and \citet{2016ApJ...833..124M}.\\

\noindent {\em P\'eclet Number. ---} The P\'eclet number is always much larger than one, with a minimum around 1000 during hydrogen burning and up to $10^{10}$ during the advanced phases. Radiative effects may still dominate at smaller scales as discussed in \citet{2015A&A...580A..61V} and they certainly play an important role during the early stages of stellar evolution. As 
mentioned in \S\ref{intro}, for most of the convective phases the evolutionary time-scale is much larger than the 
advective time-scale (Da $\sim 10^{-7}$ for hydrogen burning). Only during the later stages of evolution do these 
time-scales become comparable (Table \ref{param}; Da $>10^{-4}$).

\corr{For carbon-burning and oxygen-burning, Pe $\gg 10^6$. This is a consequence of neutrino cooling, which shortens the thermal time-scale but does not affect the radiative/conductive cooling rate. The specific entropy, $S$, obeys}

\corr{\begin{equation}
dS/dt = \partial S/\partial t + (1/\rho) \nabla { \cdot \boldsymbol{v}  } \rho S = \epsilon/T - (1/\rho T) \nabla { \cdot \boldsymbol{F_{rad}}, }
\end{equation}}

\corr{where $\rho,\boldsymbol{v}, T$ and $\boldsymbol{F_{rad}}$ are the density, flow velocity, temperature and radiative flux, respectively. $\epsilon=\epsilon_{nuc} + \epsilon_\nu$ is the net heating from nuclear burning and neutrino cooling. If $\epsilon = 0$, Rayleigh's criterion for convection may be derived \citep{1973Sci...180.1356T}. If $\boldsymbol{ F_{rad}}=0$ then the condition for simmering convection during a thermal runaway may be found \citep{1968Natur.219.1344A}.}\\

\noindent {\em Bulk Richardson Number. ---} Another important result relates to the bulk Richardson number which \corr{is a measure of } the stiffness of the convective boundary, as well as of the boundary mixing rate.  A key factor in $\rm Ri_{\rm B}$ is the buoyancy jump at the boundary (Eq. \ref{buoy}) which has contributions from both entropy and mean molecular weight ($\mu$) gradients.  At the start of burning, the \corr{thermal component of the entropy gradient} dominates.  However, as nuclear burning proceeds, the $\mu$ gradient increases and starts to dominate over the thermal component. Even during the hydrogen burning phase where the convective core continuously recedes, the $\mu$ gradient ultimately dominates over the thermal component. 

\corr{ The Richardson number\footnote{\corr{ Here we use the bulk Richardson number to denote a global measure of the stiffness of boundaries,  but do not preclude the possibility that  other varieties of Richardson number may eventually prove advantageous \citep[e.g.][]{2015ApJ...809...30A}.}} measures the ratio of potential energy from stable stratification to the turbulent kinetic energy (TKE) at the boundary, and so provides an asymptote for entrainment solutions; mixing is limited  by the energy  available. The actual rate of entrainment depends also upon  the 
effectiveness with which that energy is deposited in the stable layer rather than being advected back into the convective region (which may be related to the P\'eclet number). DNS simulations (e.g.,  \cite{2013JFM...732..150J}) typically use Pe $\sim 1$,  appropriate for air and not far from Pe $\sim7$ which may be more appropriate for water. Experiments usually have comparable P\'eclet numbers. }

\par During the advanced burning stages (C, Ne, O, and Si burning), the convective core grows during most of the stage and the boundary becomes 
`stiffer' as  $\mu$ gradients increase. As the end of the burning stage is approached, the convective regions recede and 
the boundary stiffness decreases as  the $\mu$ gradient is weakened. 

\par {We compared the bulk Richardson number between different phases and found in general that the boundary was 
at its `stiffest' during the maximum mass extent of the convective regions, and `softest' at the very end of each burning stage. 
The values we estimated for Ri$_{\rm B}$ for core carbon and oxygen burning (see Table \ref{param}) agree well with the trend described above. 
The evolution of Ri$_{\rm B}$ for the other core burning stages, however,  does not necessarily follow the same trend. This is partly due to the fact that it is not straightforward to estimate Ri$_{\rm B}$ from a 1D model.  In particular, it is not easy to define the integration length, $\Delta r$ to be used in calculating the buoyancy jump defined in Eq. \ref{buoy} \citep[see][for additional details]{2016PhyS...91c4006C}.}

\par Ri$_{\rm B}$, and thus the character of stellar convective boundaries, can be expected to vary significantly during the course of stellar evolution. \corr{Therefore, developing a convective boundary mixing model that incorporates this information would be a major advancement over most of the models currently in use.} 

\par Finally, the lower boundary of the convective shells are consistently found to be stiffer than the upper boundary.  This has important implications for astrophysical phenomena that involve CBM at the lower boundaries of convective shells. 
For example, the onset of novae \citep{2013ApJ...762....8D}, and flame front propagation in S-AGB stars \citep{2013ApJ...772...37D} \corr{which can change the model from being an electron-capture supernova progenitor to a core-collapse supernova progenitor \citep{2013ApJ...772..150J}.}

\subsection{Initial Model for 3D Hydrodynamic Simulations}

\par {We focus in this study on the second carbon burning shell of the 
15$\,\rm M_\odot$ star shown in Fig. \ref{kipp}.} Choosing the carbon shell as opposed to the core allows us to 
study two physically distinct boundaries rather than one. 

\begin{figure}
\hspace{-6mm}\includegraphics[width=0.55\textwidth,height=0.45\textwidth]{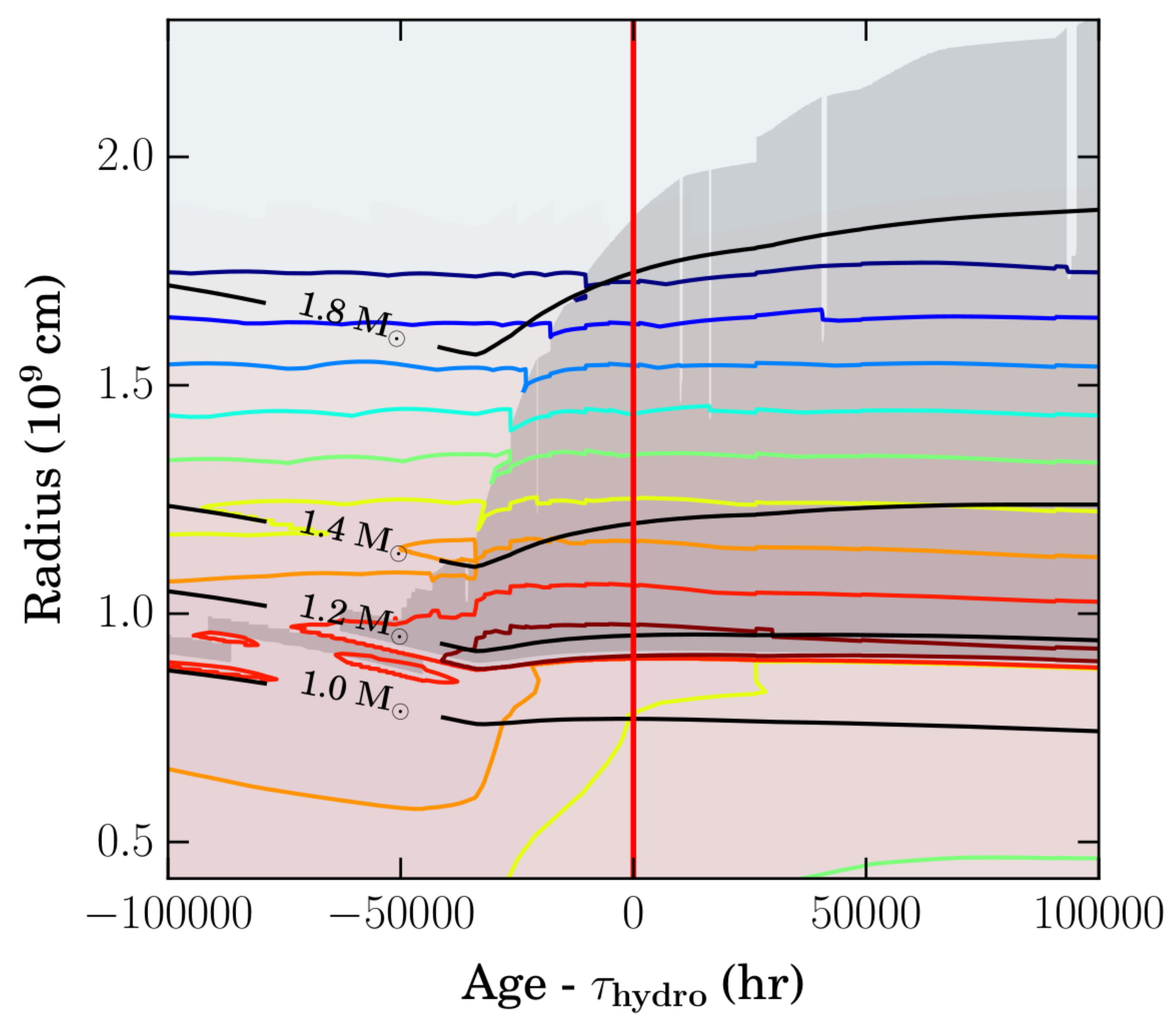}
\caption{ Convective structure evolution diagram of the 15$\,\textrm{M}_\odot$ stellar model used as initial conditions in a 3D hydrodynamics simulation friendly format. The horizontal axis is the time relative to the start of the 3D simulations 
($\tau_{\textrm{hydro}}$). The vertical axis is the radius in $10^9\,\textrm{cm}$. Mass contours in solar masses are shown by black lines and nuclear energy generation rate contours by 
coloured lines, dark red corresponds to $10^9\,$erg$\,$g$^{-1}\,$s$^{-1}$, the remaining colours decrease by one order of magnitude. Blue and pink shading represent regions of negative and positive net energy generation, respectively. 
Grey shaded areas correspond to convective regions. The vertical red bar indicates the start
time and radial extent of the hydrodynamical 3D simulation. The physical time of the simulation is on the order of 1 hour, still much shorter than the time-scale of this plot.}
\label{kipp_hydro}
\end{figure}

Figure \ref{kipp_hydro} presents a Kippenhahn diagram for the carbon shell region. The vertical red bar in this figure
shows the time at which the simulations start as well as the vertical extent of the computational domain used.
The horizontal axis shows the age of the star relative to its age at the start of the 3D
hydrodynamic simulations. We can see in Fig. \ref{kipp_hydro} that the 3D 
simulations correspond to the initial phase of the carbon burning shell, during which the convective shell 
grows in mass in the 1D model. The physical time of the 3D simulations, however, is on the order of hours, much 
shorter than the  time-scale on the horizontal axis. Furthermore, the bottom of the convective shell is 
stable (horizontal mass contour for 1.2\,$M_\odot$).  Thus, we do not expect strong structural 
re-arrangements (not considered in the 3D 
simulations as we are using a constant gravity, see \S\ref{sec:init-conditions}) to occur over the time-scale of the 3D simulations.
The mass extent of the computational domain is $0.4\, \rm M_\odot<\textrm{M}<2.1\, \rm M_\odot$ and as can be seen in Fig. \ref{kipp_hydro}  
the domain contains a stable radiative zone on both sides of the convective shell.

\section{3D Hydrodynamic Simulations}\label{method}

\subsection{The Physical Model}

We compute 3D hydrodynamic simulations using the \textsc{prompi} code \citep{2007ApJ...667..448M}. \textsc{Prompi} 
is an MPI-parallelised, finite-volume, Eulerian code derived from the legacy astrophysics code \textsc{prometheus} 
\citep{1989nuas.conf..100F}, which uses the piecewise parabolic method (PPM) of \citet{1984JCoPh..54..174C}. 
The base hydrodynamics solver can be complemented by several micro-physics prescriptions: the Helmholtz EOS of 
\cite{2000ApJS..126..501T}; an arbitrary nuclear reaction network; self-gravity in the Cowling 
approximation \citep[e.\,g. pg. 86 of][]{2000itss.book.....P} relevant for deep interiors; multi-species advection; and 
radiative diffusion (although neglected in these simulations). 

\par \textsc{prompi} solves the Euler equations (inviscid approximation), given by:

\begin{align}
\frac{\partial\rho}{\partial t} + \boldsymbol{\nabla}\cdot(\rho \,\boldsymbol{v}) &= 0;\\
\rho\,\frac{\partial \boldsymbol{v}}{\partial t} + \rho\,\boldsymbol{v}\cdot\boldsymbol{\nabla}\boldsymbol{v} &= -\boldsymbol{\nabla}p +\rho\,\mathbf{g};\\
\rho\,\frac{\partial E_t}{\partial t} + \rho\,\boldsymbol{v}\cdot\boldsymbol{\nabla}E_t + \boldsymbol{\nabla}\cdot(p\,\boldsymbol{v})&=\rho\,\boldsymbol{v}\cdot\mathbf{g}+\rho(\epsilon_{nuc}+\epsilon_\nu);\\
\rho\,\frac{\partial X_i}{\partial t} + \rho\,\boldsymbol{v}\cdot\boldsymbol{\nabla}X_i &= R_i,
\end{align}

\noindent where $p$ is the pressure, $\mathbf{g}$ the gravitational 
acceleration, $E_t$ the total energy, $X_i$ the mass fraction of nuclear species $i$ and $R_i$ the rate of change of nuclear species $i$.

\par While there is evidence that magnetic fields will be generated in deep interior convection 
\citep[e.\,g.][]{2004PhRvL..92n4501B} and that
rotational instabilities \citep[e.\,g.][]{2013A&A...553A...1M}  may play an important role in shaping convection, we focus purely on
the hydrodynamic aspects in the current study, which remains a problem of significant complexity with many outstanding issues.

\par Energy generation during carbon burning proceeds mainly via fusion of two $^{12}$C nuclei. For stellar 
conditions, considering only the main exit channels ($\alpha$ and $p$) will result in no significant errors 
\citep{1996snih.book.....A}. The $n$ exit channel 
branching ratio is only $b_n=0.02$, so for this study we only consider energy generation due to the $\alpha$ and $p$ channels. We estimated the carbon burning energy generation rate in our 3D simulations with a 
slightly modified version of the parameterisation given by \citet{1986nce..conf.....A} and \citet{2009pfer.book.....M}:\\
\vspace{-4mm}
\begin{equation}
\epsilon_{^{12}C} \sim 4.8\times10^{18}\;Y_{12}^2\; \rho\; \lambda_{12,12},
\end{equation}

\noindent where $Y_{12} = X_{^{12}\textrm{C}}/12$, $\lambda_{12,12} = 5.2\times10^{-11}\, T_9\hspace{0.05mm}^{30}$ and $T_9=T/10^9$.

\par This simplification to the nuclear physics allows us to represent the stellar material using only three compositional quantities: the average 
atomic mass $\bar{A}$, average atomic number  $\bar{Z}$, and the carbon abundance $X_{^{12}\textrm{C}}$. The mass and charge are 
required for the EOS and to represent the mean properties of all other species besides $^{12}$C. \corr{ Thus the composition is an active scalar, and coupled to the flow through the EOS and mixing.}
A further simplification is that the change of
$^{12}$C due to nuclear burning was ignored because of its negligible rate of change relative to advective mixing over such short time-scales
(i.\,e. the carbon shell is
characterised by a very small  Damk\"{o}hler number, Da $\sim 10^{-4}$, see Table \ref{param}). The key important 
feature retained with this prescription of the nuclear burning is the interaction and feedback between the 
nuclear burning and hydrodynamic mixing, while keeping computational costs to a minimum.

\par Cooling via neutrino losses is parameterised using the analytical formula provided by \citet{1967ApJ...150..979B} which includes all of the 
relevant processes: pair creation reactions; Compton scattering; and plasma neutrino reactions. The cooling is essentially constant \corr{ over the simulation time }and its details are not important for our purposes.

\subsection{The Computational Domain}\label{comp-dom}

\par \corr{ Approximations are necessary to simulate a 
meaningful physical time.} In this study, we follow the ``box-in-star'' approach \citep{2016RPPh...79j2901A} and we use a Cartesian coordinate system and a plane-parallel geometry. 
We evolve the model with time-steps determined by the Courant condition, using a Courant factor of 0.8. \corr{ Our computational domain represents a convective region of thickness, $t$, bounded either side by radiative regions of thickness, $t/2$. The aspect ratio of the convective zone is therefore 2:1 (width:height), and so a plane-parallel approximation is not ideal and is the first major simplification of our set-up.} We made this choice to allow us to ease the difficult Courant time-scale condition at the inner boundary of the grid allowing for longer run-times, \corr{ as well as} better resolution near convective boundaries. \corr{ Direct comparison with the oxygen burning simulations, which use a spherical grid, suggest that no significant error results.}

\begin{figure}
\hspace{-6mm}\includegraphics[width=0.55\textwidth,height=0.45\textwidth]{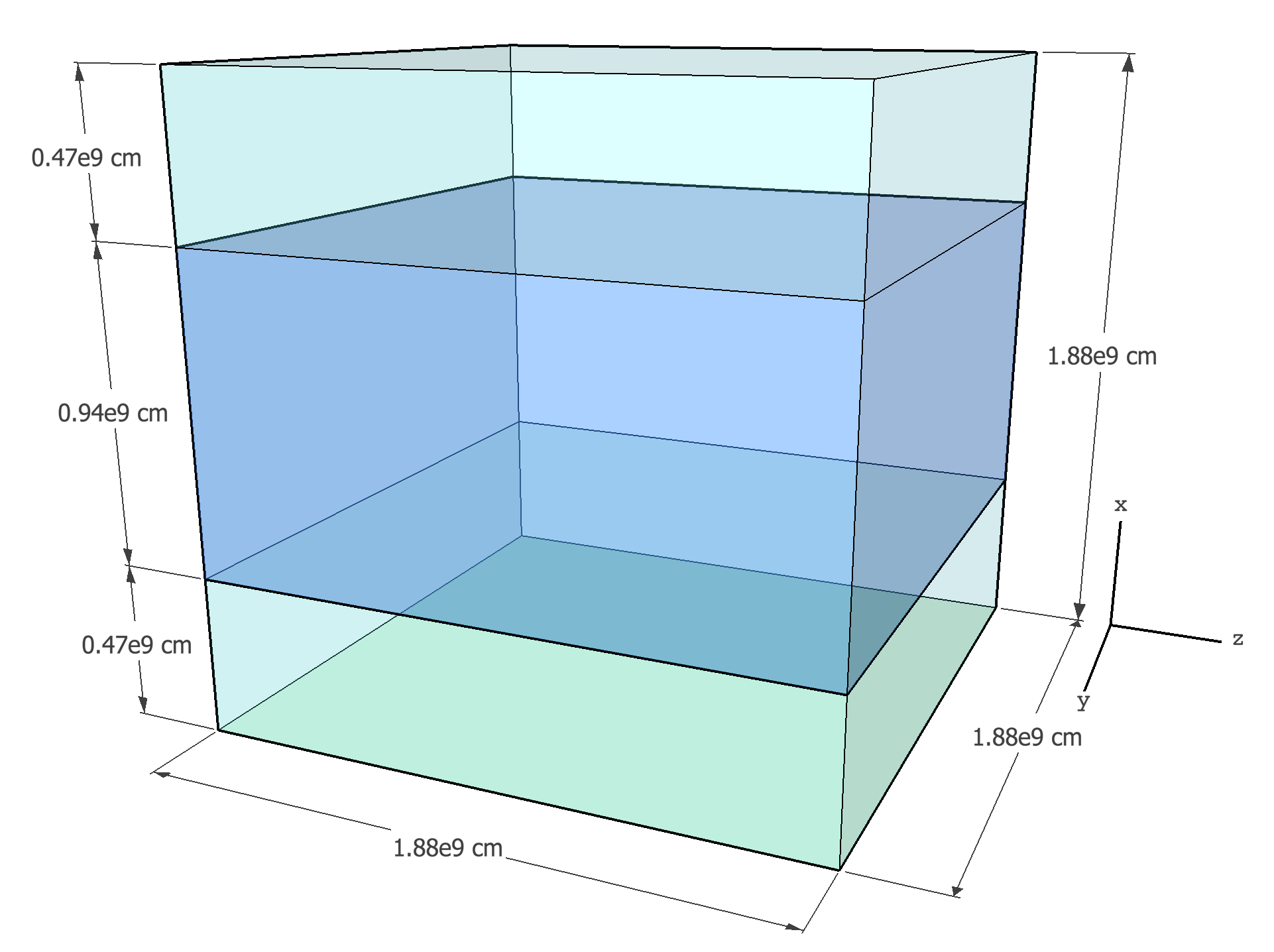}
\caption{\corr{The geometry of the computational domain. Gravity is aligned with the $x$-axis. The blue region depicts the approximate location of the convectively unstable layer at the start of the simulation while the surrounding green volumes depict the locations of the bounding stably stratified layers. \label{comp-domain}}}
 \end{figure}
 
\par In order to study the complete convective region, and also stable region dynamics (such as wave propagation), we chose to include 
the entire convection zone and portions of the adjacent stable regions. The radial 
extent of the domain in relation to the stellar model initial conditions is illustrated in Fig. \ref{kipp_hydro} by the vertical red bar.
\corr{ The computational domain extends in the vertical (x) direction 
from $0.42\times10^9\,\textrm{cm}$  to $2.30\times10^9\,\textrm{cm}$, and in the two horizontal directions (y and z) 
from $0$ to $1.88\times10^9\,\textrm{cm}$,} \corr{see Fig.~\ref{comp-domain}.}

\par \corr{We found that the aspect ratio for the convective zone of 2:1 was the required minimum for unrestricted circulation of turbulent fluid elements. The radial extent of the computational domain represents \corr{ $4\times10^{-5}$ of} the total radius of the star, which is $4.6\times10^{13}$\,cm. At the chosen evolutionary stage, the shell is expanding, as can be seen in Fig. \ref{kipp_hydro}, and the luminosity is driven by a peak in nuclear energy generation of $\sim 10^9\,\textrm{erg}\,\textrm{g}^{-1}\,\textrm{s}^{-1}$
at x $\sim0.9\times10^9\,\textrm{cm}$.}

\par The computational domain uses reflective boundary conditions in the vertical direction and periodic boundary conditions in the two 
horizontal directions.  Although the material in the radiative regions is stable against convection it has oscillatory g-mode motions excited by 
the adjacent convection zone. In order to mimic the propagation of these waves out of the domain, \corr{ we employ a damping region that extends radially between a radius of $0.6\times10^9$\,cm 
and the lower domain boundary at $0.42\times10^9$\,cm. The damping region covers the full horizontal extent of the computational domain in between these radii. Within this region all velocity components are reduced by a common damping factor, $f$, resulting in damped velocities over the damping region, $\boldsymbol{v_{d}}=f\boldsymbol{v}$. The damping factor is defined as}

\corr{\begin{equation}
f=\left(1+\delta t\, \omega f_{d}\right)^{-1},
\end{equation}}

\corr{where $\delta t$ is the time step of the simulation, $\omega=0.01$ is the damping frequency and is a free parameter chosen to correspond to a small fraction of the convective turnover. $f_{d}=0.5\;(\textrm{cos}\left(\pi r/r_0\right)+1)$, where $r$ is the radial position in the vertical direction and $r_0 = 0.6\times10^9$\,cm is the edge of the damping region in the vertical direction. Using this damping function, $f_d =0$ at $r=r_0$, where the damping region starts. This ensures a smooth transition between the non-damped and damped region. }

\par To test the dependence of our results on numerical resolution we simulated the carbon shell at four different resolutions. These models are named according to their resolution: \textsf{lrez} - 128$^3$, \textsf{mrez} - 256$^3$, \textsf{hrez} - 512$^3$ and 
\textsf{vhrez} - 1024$^3$.

\corr{Whether a computed flow will exhibit turbulence depends on the spatial and temporal discretisation that is used. In the following we explore heuristically the 3D modelling of turbulence on a discrete grid.}

\subsubsection{\corr{Spatial zoning considerations}}\label{spatial_turb}   

\corr{A useful dimensionless number for determining the \corr{degree} of turbulence in a simulation is the effective or {\em numerical} Reynolds number, a discrete analogue of the Reynolds number. It can be defined using the following arguments.}

\corr{\citet{1941DoSSR..30..301K} showed that the rate of energy dissipation at any length-scale, $\lambda$ (between the inertial range and Kolmogorov scale), is given by $\epsilon_\lambda\sim v_\lambda^3/\lambda$, where $v_\lambda$ is the flow velocity at that scale. This relation can be applied at the extreme scales of the simulation i.e. at the integral scale and the grid scale to give}

\corr{\begin{equation}\label{eps_scales}
\epsilon_{\ell}=\frac{v_{rms}^3}{\ell}\;\;\;\textrm{and}\;\;\; \epsilon_{\Delta x}=\frac{\Delta u^3}{\Delta\, x},\;\;\;\textrm{respectively},
\end{equation}}

\corr{where $\Delta u$ is the flow velocity across a grid cell. This velocity can also be used to define an effective numerical viscosity at the grid scale}

\corr{\begin{equation}\label{nu_eff}
\nu_{\rm eff}=\Delta u \Delta x.
\end{equation}}

\corr{For a turbulent system within a statistically steady state \citet{1962JFM....13...82K} showed that the rate of energy dissipation is equal at all scales. Applying this equality to Eq. \ref{eps_scales} yields (with the use of Eq. \ref{nu_eff})}

\corr{\begin{equation}
\nu_{\rm eff}=v_{rms}\ell \left(\frac{\Delta x}{\ell}\right)^{4/3}.
\end{equation}}

\corr{Therefore the effective Reynolds number can be expressed as}

\corr{\begin{equation}\label{re_eff}
\textrm{Re}_{\rm eff}=\left(\frac{\ell}{\Delta x}\right)^{4/3} \sim N_x^{\,4/3},
\end{equation}}

\corr{where $N_x$ is the number of grid points in the vertical direction. In these simulations this is a slight over-estimate as in the vertical direction only half of the grid points represent the convective region.}

\corr{Within the ILES paradigm the effective Reynolds number is therefore limited by the momentum diffusivity\footnote{\corr{The actual numerical dissipation of the PPM method is highly
complex and non-linear \citep{2000JCoPh.158..225S}; the highest resolution simulations presented here seem
to capture the effective dissipation accurately.}} at the grid scale (Eq. \ref{nu_eff}), and as demonstrated by Eq. \ref{re_eff} it is the choice of spatial zoning that sets an upper limit on the degree of turbulence. The effective Reynolds numbers of our simulations ($N_x=128 - 1024$) therefore range from around 650 to 10$^4$, suggesting that we are within the turbulent regime (Re$_{\rm eff} \gtrsim 1000$) for the finer grids\footnote{\corr{This is supported by visual comparison of our simulations with experimental data \citep[e.g.][]{1982STIA...8236549V}.}}. }

\subsubsection{\corr{Time-scale considerations}}

\begin{figure*}
\minipage{0.475\textwidth}
 \includegraphics[width=\linewidth]{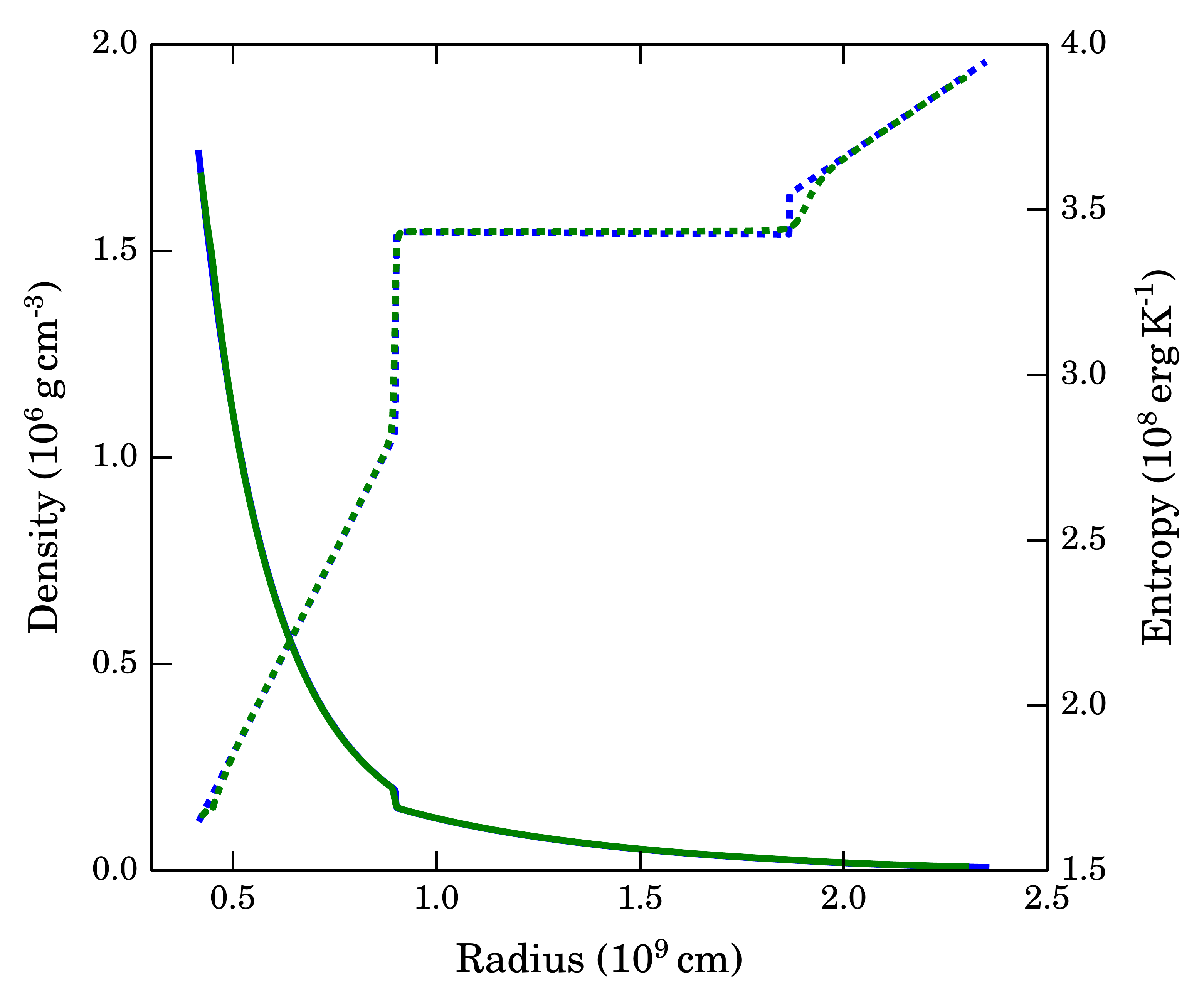}
\endminipage\hfill
\minipage{0.475\textwidth}
 \includegraphics[width=\linewidth]{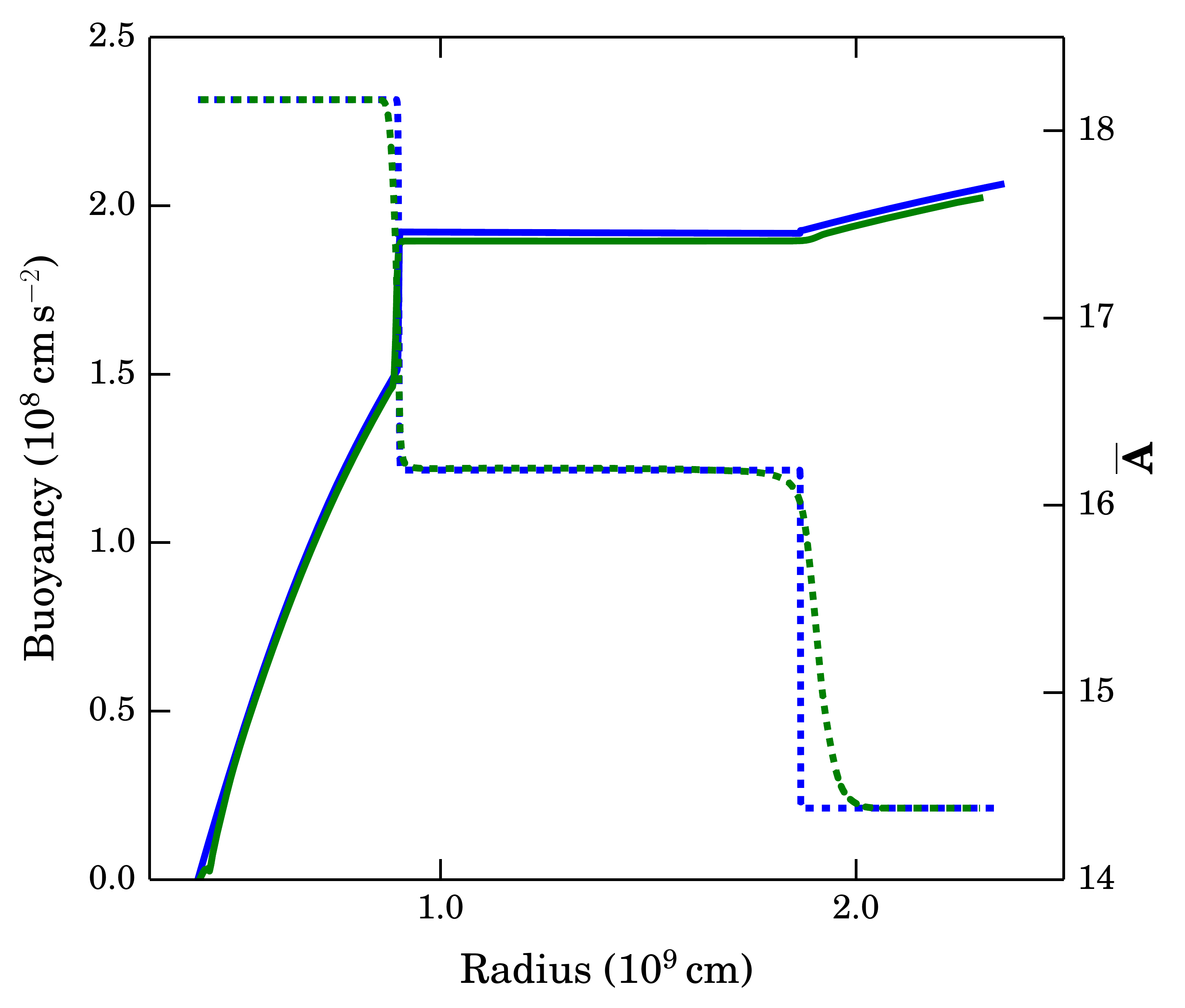}
\endminipage\hfill
\caption{\textsf{Left:} Initial radial density (solid) and entropy (dashed) profiles. \textsf{Right:} Initial radial buoyancy (solid) and composition (dashed) profiles. One dimensional stellar evolution 
profiles calculated using \textsc{genec} (blue) are compared with the same profiles integrated and mapped onto the Eulerian Cartesian grid in \textsc{prompi} (green).}
\label{1d3d}
\end{figure*} 

\corr{The convective turnover time, $\tau_c$ (twice the transit time), is the time needed to set up the turbulent velocity field \citep{2007ApJ...667..448M}, following the initial perturbations in temperature and density. \corr{Therefore, {\em the convective turnover time is the minimum time-scale required for simulating turbulence}}. For carbon burning the turnover time is $\tau_c \sim 7\times10^3$\,s. The maximum time step size allowed by the explicit hydrodynamic solver is $\Delta t_{max} = \Delta x/c_s$, where the sound speed is approximately $c_s \sim 4.5\times10^8$\,cm\,s$^{-1}$. 
\corr{Therefore, the minimum number of time-steps needed to simulate a convective turnover time is $N_{\Delta t} \sim N_x/Ma$ for Mach number $Ma$.} For the \textsf{hrez} zoning ($N_x= 512$), this equates to $\sim8000$\,s which would exceed the available computer resources by a factor of $\sim 500$ per simulation.}

\corr{Hence, as one may guess intuitively, the modelling of smaller velocities requires more time steps. One option to overcome this issue is to scale the velocity up by scaling the nuclear energy generation rate. Scaling the burning rate by a factor of 1000 reduces the convective turnover time to $\tau_c\sim 270$\,s, and the minimum number of time-steps required to establish a turbulent flow decreases to $N_{\Delta t}\sim 270N_x$, which for the \textsf{hrez} zoning is around 750\,s, which is comfortably attainable given the available computational resources.}

\subsubsection{\corr{Boosting factor}}

\corr{A boosting factor of 10$^3$ for the nuclear energy generation rate was chosen in order for the simulations to match the turbulent driving observed in oxygen shell burning simulations \citep[$\sim10^{12}$\,erg\,g$^{-1}$\,s$^{-1}$;][]{2007ApJ...667..448M}.}

In such simulations there is no need to worry about the effect that such a boost in driving will have on the thermal diffusion in the model as it can be safely ignored \corrdre{in the bulk of the convective zone.} This is because thermal diffusivity is negligible in comparison to the loss of heat through escaping neutrinos produced in the plasma \citep[][pg. 284 - 292]{1996snih.book.....A}, and so thermal diffusion implicitly only becomes important at the sub-grid scale \corrdre{\citep[see also discussion in][]{2015A&A...580A..61V}. Although future studies are needed to confirm the P\'{e}clet number in the boundary layers, \citet{2015ApJ...809...30A} argue that thermal diffusivity is also very small in the boundary regions of the oxygen burning shell, which would also apply to our boosted carbon shell. They show that a large P\'{e}clet number leads to an adiabatic expansion of the convective boundary.}


\corr{This boosting of the driving luminosity does not have any dynamical effect on the shell structure, given the short physical time-scales of the simulations. The convective velocities and boundary mixing rates will be increased though, compared to the astrophysical scenario being modelled. A key advantage to this approach is that more convective turnovers can be simulated for the given physical time that is being modelled, 
but it does highlight an important sensitivity of the hydrodynamic flow to the numerical set-up.  Additionally, as the nuclear luminosity has been boosted the neutrino losses contribute negligibly to the thermal evolution of the model.}

\subsection{Initial Conditions and Runtime Parameters}\label{sec:init-conditions}

\par The initial vertical extent of the convective region ($0.90\times 
10^9\,\textrm{cm} \lesssim \textrm{x} \lesssim 1.87 \times 10^9\,\textrm{cm}$) can be seen through 
the entropy, buoyancy and composition profiles in Fig. \ref{1d3d}. The convective region is apparent 
through the homogeneity of these quantities due to strong mixing, while the boundaries are defined by sharp jumps. 

\par An initial hydrostatic structure in \textsc{prompi} was reconstructed from the entropy, composition and gravitational acceleration profiles taken from the \textsc{genec} 1D model. Stellar models do not have regularly spaced mesh points in the radial direction given the fact that they use a Lagrangian method and so the spatial resolution is sometimes coarse, especially at convective boundaries. For this reason, the 1D \textsc{genec} profiles of the entropy ($s$), average atomic mass ($\bar{A}$) and average atomic number ($\bar{Z}$) were first remapped onto a finer grid mesh before linearly interpolating onto the Eulerian grid in \textsc{prompi}. The details of this re-mapping can be found in Appendix \ref{app}.

\par There is no nuclear burning network in this model, in the sense that we do not follow the depletion of $^{12}$C through nuclear burning, but only through mixing. The abundance variables $\bar{A}$, $\bar{Z}$ and $X_{^{12}C}$ are somewhat redundant though, as the electron fraction $Y_e=\bar{Z}/\bar{A}$ does not change. 

\par To ensure the model is in hydrostatic equilibrium, the density $\rho\,(s,p,\bar{A},\bar{Z})$ was integrated along the new 
radial grid according to:

\begin{table}
\caption{Summary of simulation properties. $N_{xyz}$: Total number of zones in the computational domain ($N_x\times N_y\times N_z$), $\tau_{\rm sim}$: simulated physical time (s), $v_{\rm rms}$: global RMS convective velocity (cm\,s$^{-1}$), $\tau_c$: convective turnover time (s), $\textrm{Ri}_\textrm{B}$: bulk Richardson number (values in brackets are representative of the lower convective boundary region), Ma: Mach number.}
\begin{tabulary}{0.5\textwidth}{l || c c c c}
\hline \hline\\
 & \textsf{lrez} & \textsf{mrez} & \textsf{hrez} & \textsf{vhrez} \\\\
\hline \hline\\
\textsf{N$\boldsymbol{_{xyz}}$}  &  128$^3$  &  256$^3$  &  512$^3$  &  1,024$^3$ \\\\
\textsf{$\boldsymbol{\tau_{sim}}$}  &  3,213  &  3,062  &  2,841  &  986 \\\\
\textsf{$\boldsymbol{v_{rms}}$} & 3.76$\times10^6$ & 4.36$\times10^6$ & 4.34$\times10^6$ & 3.93$\times10^6$ \\\\
\textsf{$\boldsymbol{\tau_c}$} & 554 & 474 & 471 & 513 \\\\
\textsf{$\boldsymbol{\textrm{Ri}_\textrm{B}}$} & 29 (370) & 21 (259) & 20 (251) & 23 (299) \\\\
\textsf{Ma} & 0.0152 & 0.0176 & 0.0175 & 0.0159 \\\\
\hline \hline
\end{tabulary}
\label{tab1}
\end{table}

\begin{align}\label{dddr}
\begin{split}
\frac{\partial \rho}{\partial r} =\, & \frac{ds}{dr} \left(\frac{\partial \rho}{\partial s}\right)_{p, \bar{A}, \bar{Z}} + \frac{dp}{dr} \left(\frac{\partial \rho}{\partial p}\right)_{s, \bar{A}, \bar{Z}} + \frac{d\bar{A}}{dr} \left(\frac{\partial \rho}{\partial \bar{A}}\right)_{s, p, \bar{Z}}\\ 
&+ \frac{d\bar{Z}}{dr} \left(\frac{\partial \rho}{\partial \bar{Z}}\right)_{s, p, \bar{A}},
\end{split}
\end{align}

\noindent the second term is simplified by enforcing hydrostatic equilibrium to within a tolerance of $10^{-10}$, given by:

\begin{equation}\label{hse}
\frac{dp}{dr} = -\rho g.
\end{equation}

For our plane-parallel geometry set-up, the gravitational acceleration was parameterised by a function of the form $g(r)=A/r$, with constant $A=1.5\times10^{17}\,\textrm{cm}^2\,\textrm{s}^{-2}$. 
The total derivatives $ds/dr$, 
$d\bar{A}/dr$ and $d\bar{Z}/dr$ were calculated from the fitted profiles introduced earlier. 
The partial derivatives $\partial\rho/\partial s$,\; $\partial\rho/\partial p$,\; 
$\partial\rho/\partial \bar{A}$ and $\partial\rho/\partial \bar{Z}$ were calculated using the Helmholtz EOS \citep{1999ApJS..125..277T,2000ApJS..126..501T}. 
Figure \ref{1d3d} shows the density, entropy, buoyancy and average atomic mass profiles for the stellar model initial conditions, 
and the corresponding initial profiles that were mapped onto the Eulerian grid in \textsc{prompi}.
\par Simulation time is typically measured in convective turnovers, $\tau_{c} = 2\,\ell_{c}/v_{rms}$ where $\ell_{c}$ is the height of the convective 
region and $v_{rms}$ is the global convective velocity, 
$v_{rms}=\sqrt{\overline{\left<v_x^{\,2}\right>}-\overline{\left<v_x\right>}^{\,2}}$ (see Appendix \ref{app-tke} for a 
description of this notation). Our simulations typically span 3 to 4 turnovers, following an initial transient phase of around 1,000\,s.

\par \corr{Convection} is seeded in the hydrodynamic models \corr{through} random perturbations in temperature and density in the same manner described by
\citet{2007ApJ...667..448M} who also showed that the subsequent nature of the flow was independent of these seed perturbations. For the \textsf{vhrez} model convection was not seeded through perturbations in the 1D stellar model initial conditions, but was restarted from the \textsf{hrez} model at 980\,s, this was done by duplicating each of the cells to double the resolution. 
Due to limited computational resources available for this study, the \textsf{vhrez} model was not 
simulated for enough convective turnovers in order for the temporal averaging to be statistically valid. As a result, we only included this model in part of our detailed analysis.

\begin{figure}
\includegraphics[width=0.5\textwidth]{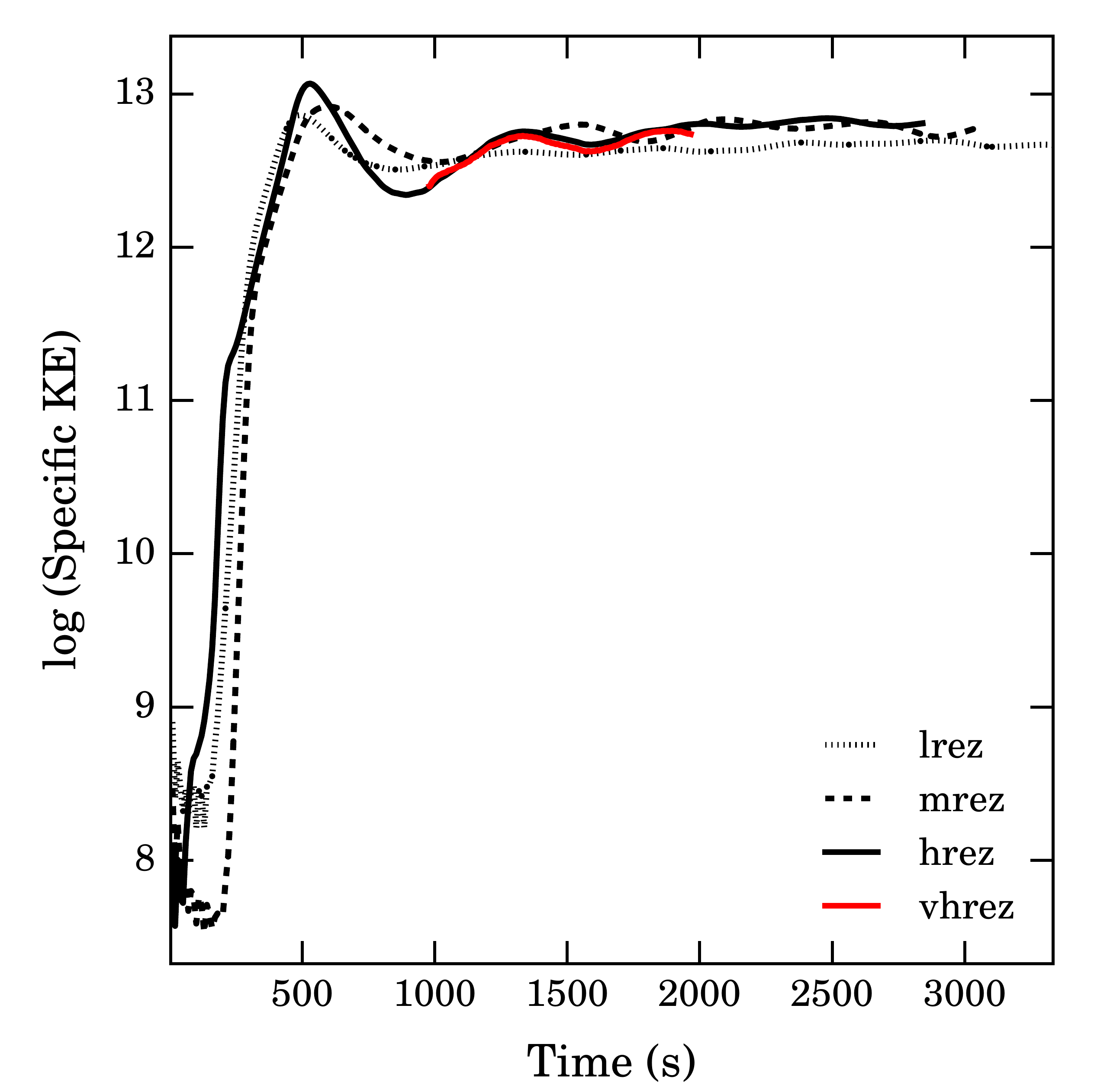}
\caption{Temporal evolution of the global specific kinetic energy: 
thin dashed - \textsf{lrez}; 
thick dashed - \textsf{mrez}; 
black solid - \textsf{hrez}; 
red solid - \textsf{vhrez}. The quasi-steady state begins in each model after approximately 1,000$\,$s, and only the \textsf{lrez} model kinetic energy appears to have a dependence on the resolution.}
\label{ek}
\end{figure} 

\section{Simulation Results}\label{results}

\par A summary of the simulation models is presented in Table \ref{tab1}, which \corr{includes the number of zones, physical time simulated, convective velocity, convective turnover time, bulk Richardson number, and convective Mach number.}

\subsection{The Onset of Convection and Time Evolution}

\begin{figure*}
\includegraphics[width=0.5\textwidth]{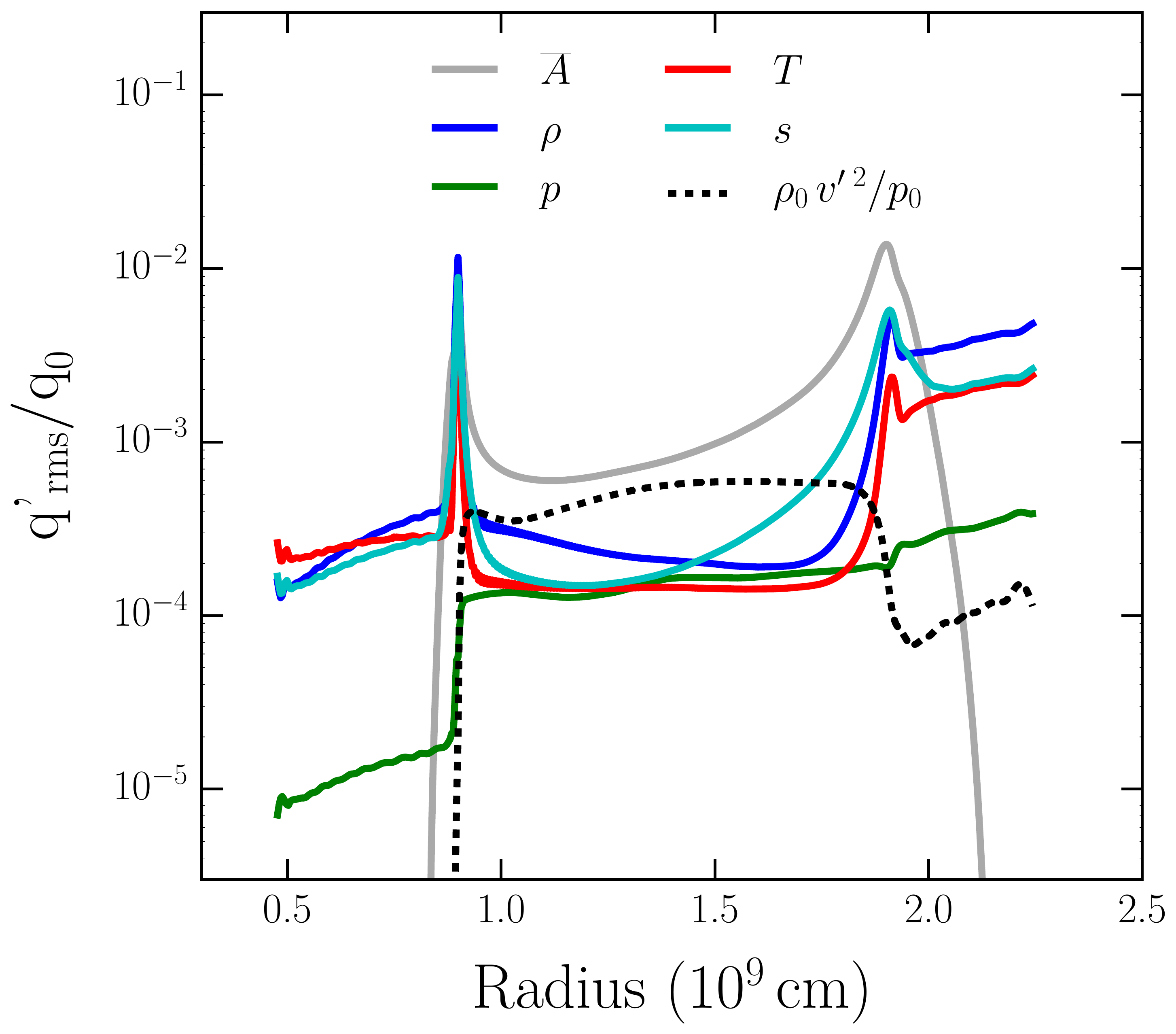} 
\includegraphics[width=0.48\textwidth,height=0.445\textwidth]{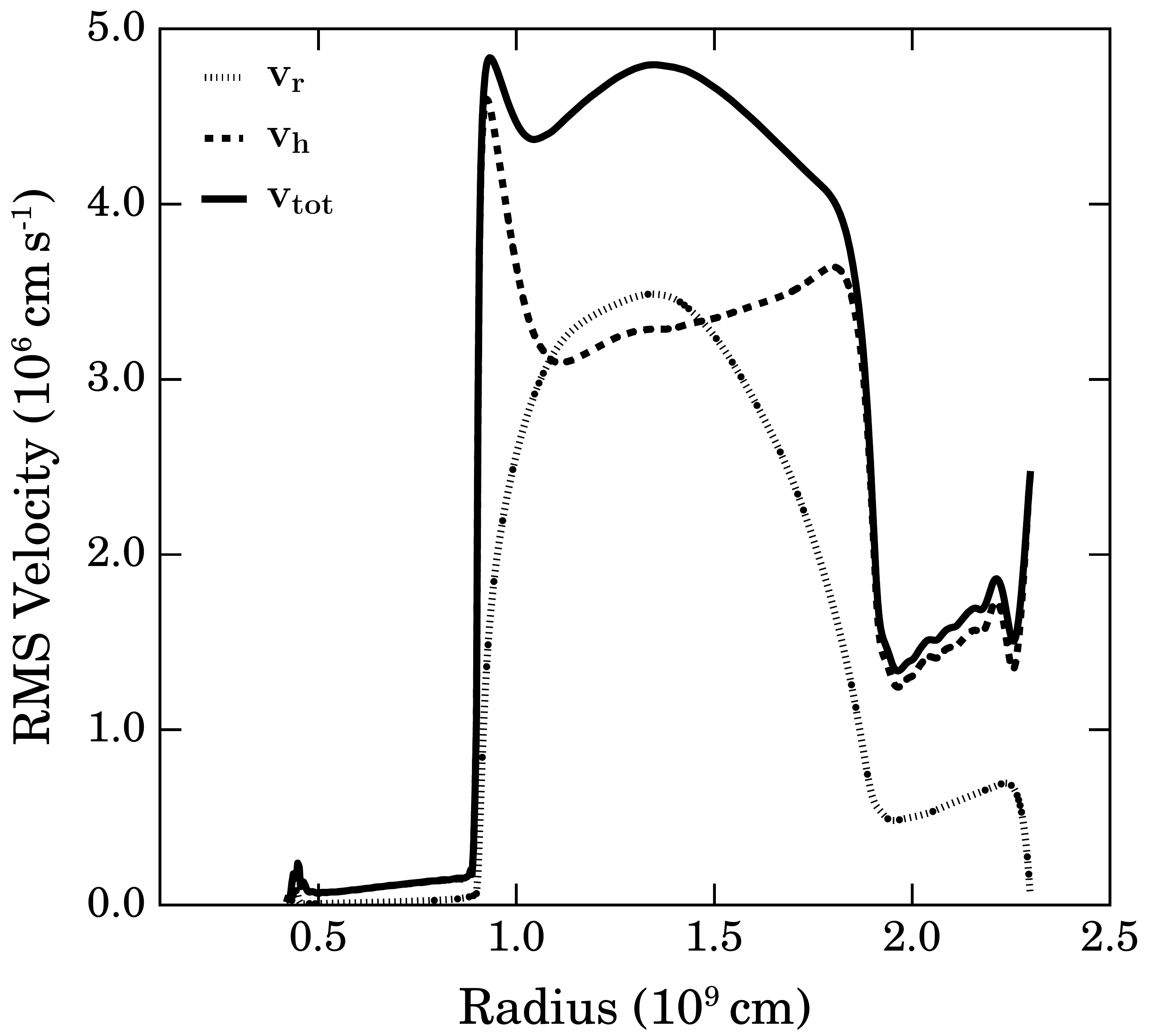} 
\caption{\textsf{Left:} Horizontally averaged RMS fluctuations of composition, density, pressure, temperature 
and entropy weighted by their average values. The dashed curve represents a pseudo-sound term. These 
fluctuations were time averaged over 4 convective turnover times of the \textsf{hrez} model. \textsf{Right:} RMS radial (thin dashed), horizontal (thick dashed) and total (solid) velocity components, time averaged over four convective  turnover times for the \textsf{hrez} model. Local maxima in the horizontal velocity indicate the approximate convective boundary locations.}
\label{urms-fluct}
\end{figure*} 

\par  The temporal evolution of the global (\corr{averaged} over the convective zone) specific kinetic energy for all of the 
models is presented in Fig. \ref{ek}. The first $\sim$1,000 seconds of evolution are characterised by an initial transient associated
with the onset of convection. By $\sim$1,250\,s all of the models settle into a quasi-steady state characterised by semi-regular
pulses in kinetic energy occurring on a time-scale of the order of a convective turnover time. These pulses are associated with the formation and eventual breakup  of semi-coherent, large-scale eddies or plumes 
that traverse a good fraction of the convection zone before dissipating, and is a phenomena that is typical
of stellar convective flow \citep{2007ApJ...667..448M,2011ApJ...733...78A,2011ApJ...741...33A,2013ApJ...769....1V,2015ApJ...809...30A}.  

\par As discussed in \S\ref{sec:init-conditions},  the evolution of the highest resolution model, \textsf{vhrez}, begins at $\sim$1,000\,s when it was restarted from model \textsf{hrez} by simply sampling the underlying flow field onto a higher resolution mesh. As is typical of turbulent flow this model relaxes in approximately one large-eddy crossing time as evidenced by the re-establishment of the TKE balance
discussed below (\S \ref{MFA}).

\par Although these simulations do not sample a large number of convective turnover times (between $\sim$2 and $\sim$6; discussed below), resolution trends are still apparent. The most prominent trend seen here is the kinetic energy peak associated with the initial transient, which
increases as the grid is refined. This is not linked to the initial seed perturbations and is most likely related to the decreased numerical dissipation at finer zoning.  

\par A similar trend can also be seen in the quasi-steady turbulent state that follows the initial 
transient. Interestingly, in this case, a resolution dependence only appears to manifest for the lowest resolution 
model, \textsf{lrez}.  This
has an overall smaller amplitude of kinetic energy as well as a much smaller variance associated with the formation 
and destruction of pulses. These properties can be naturally attributed to a higher numerical dissipation at a lower resolution,  an issue
that we return to throughout the remainder of the paper.

\subsection{Properties of the Quasi-Steady State}

\begin{figure*}
\minipage{0.33\textwidth}
 \includegraphics[width=0.975\linewidth]{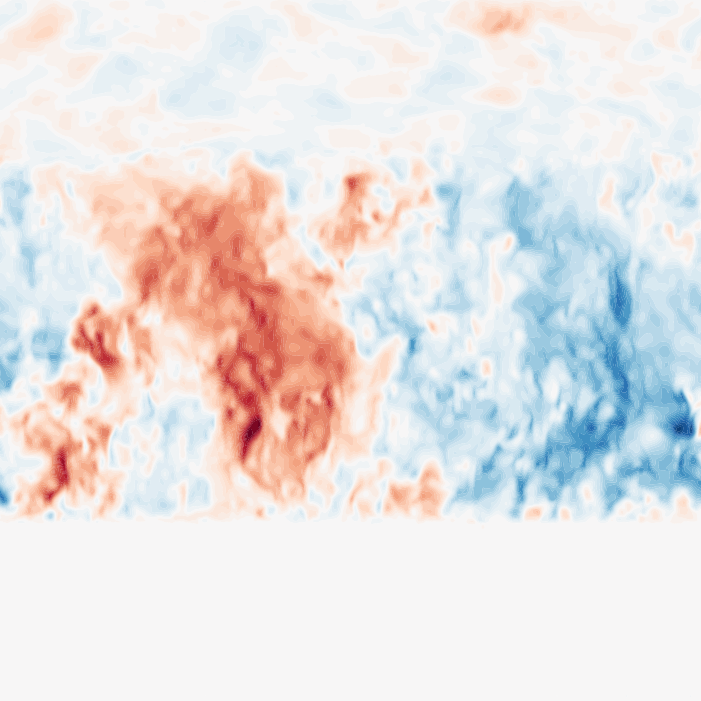}
\endminipage
\minipage{0.33\textwidth}
 \includegraphics[width=0.975\linewidth]{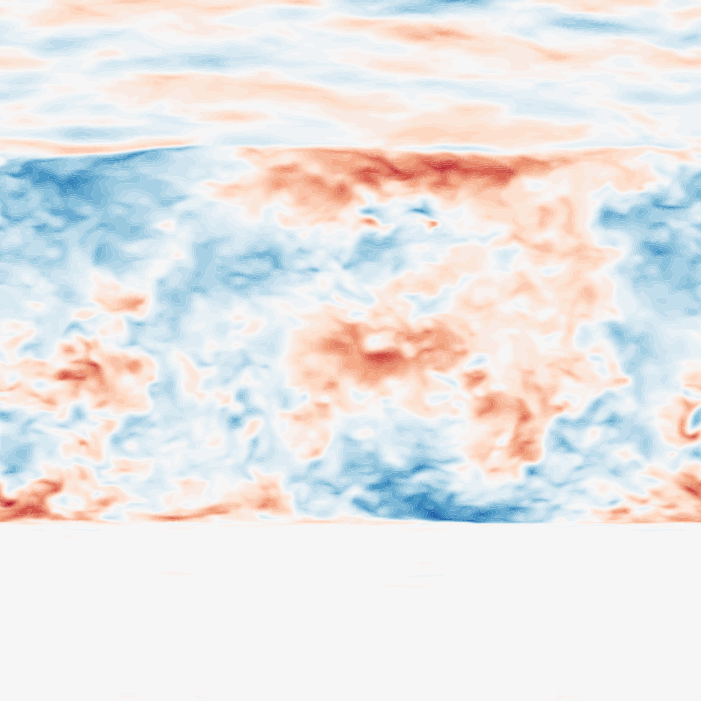}
\endminipage
\minipage{0.33\textwidth}
 \includegraphics[width=0.975\linewidth]{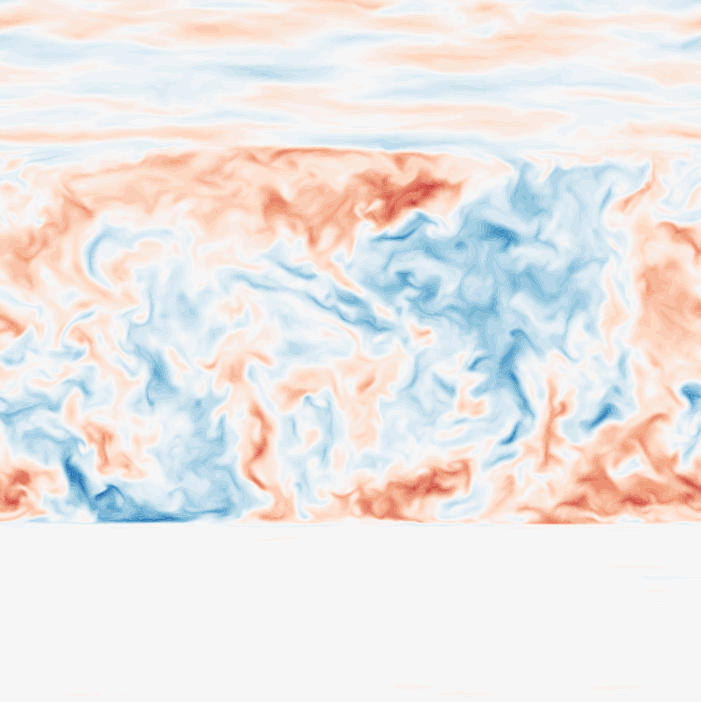}
\endminipage
\minipage{0.085\textwidth}
 \includegraphics[width=0.9\linewidth,height=4\linewidth]{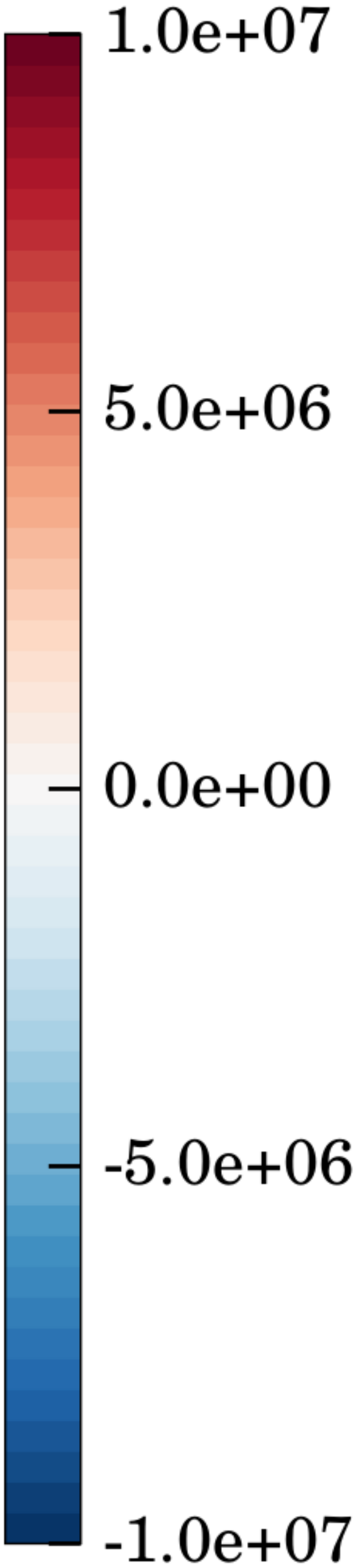}
\endminipage
\caption{Vertical 2D slice (2D plane defined by $z=0.94\times10^9\,\textrm{cm}$, where $z$ is one of the two 
horizontal 
directions in the \textsc{prompi} simulation) of velocity components, 1480s into the \textsf{hrez} 
simulation. From left to right $v_x$ (vertical component), $v_y$ and $v_z$ (horizontal 
components) are plotted. Reds are positive,
blues are negative and white represents velocities around zero.} 
\label{velx}
\end{figure*} 

\par RMS fluctuations in density, pressure, entropy, temperature and composition centred around their mean background states are shown for the \textsf{hrez} model in the left panel of Fig.~\ref{urms-fluct}. Fluctuations in the convective region are small and of a similar magnitude for all quantities except the composition. Near the convective boundary regions, the relative amplitude of the fluctuations is highest, reaching values around $1\%$ of the mean background state.

Pressure fluctuations can be grouped into a compressible and an incompressible component. The former describes the 
acoustic nature of pressure fluctuations such as when the flow turns and is compressed. The latter describes the 
advective nature of pressure perturbations due to buoyancy effects. The compressible component of the pressure 
fluctuations is proportional to a pseudo-sound term, $\rho_0 \,v'^2/p_0$, shown by the dashed line in the left of Fig. 
\ref{urms-fluct}. This term is highest in the convective region and has a magnitude similar to the square of the Mach 
number, $\sim3\times10^{-4}$.

\begin{figure}
\includegraphics[width=0.5\textwidth]{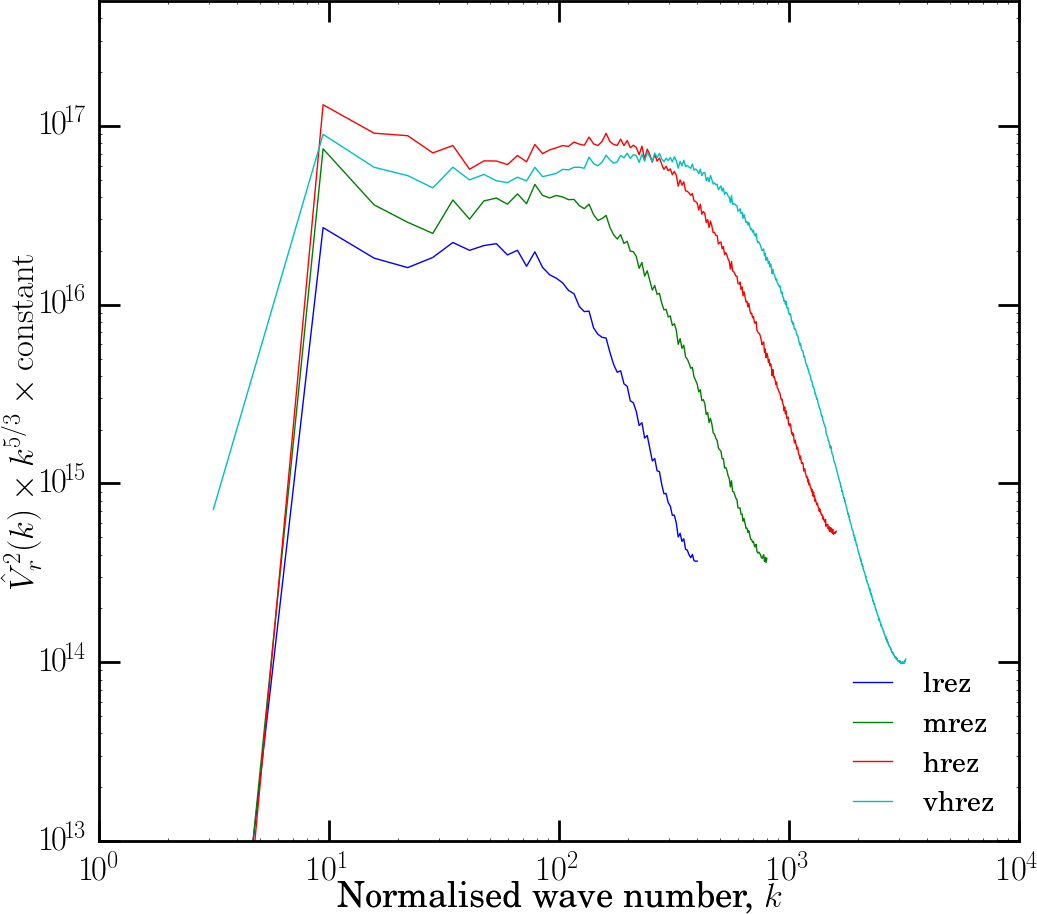}
\caption{\corr{Specific kinetic energy spectrum of the four simulations. Spectra were obtained from a 2D Fourier transform of the vertical velocity (at the mid-height of the convective region), averaged over several convective turnovers. The vertical axis corresponds to the square of the Fourier transform, scaled by a `Kolmogorov factor', $k^{5/3}$, and a constant ($N^{-1}$) to allow easier comparison between resolutions. \corrdre{This scaling highlights the sub-range of wave-numbers which obey the $k^{-5/3}$ power law \citep{1941DoSSR..30..301K}.} The horizontal axis represents the wave-number, $k$. See text for more details.}}
\label{ke_spectra}
\end{figure}

Horizontally averaged RMS velocity components for the \textsf{hrez} model are shown on the right of Fig. 
\ref{urms-fluct}. These profiles represent an average over the quasi-steady state period of the simulation, which we 
estimate to occur over four convective turnover times. The total RMS velocity reaches a maximum of around 
$4.8\times10^6$ cm$\,$s$^{-1}$ both in the centre of the convective region ($x \sim 1.4\times10^9\,$cm) and also near 
the lower convective boundary ($x \sim 0.9\times10^9\,$cm). Contributions to the total velocity are dominated by the 
radial velocity over the \corr{central part} of the convective region, while close to the convective boundaries the horizontal velocity 
($v_h=\sqrt{v_y^2+v_z^2}$\,) is the largest component. The local maxima in horizontal velocities correspond to the radial 
deceleration and eventual turning of the flow near the convective boundaries. Such features are typical of shallow convective regions and are similarly reported in simulations of the oxygen burning shell by \citet{2007ApJ...667..448M} and \citet{2017MNRAS.465.2991J}, see their figs. 
6 and 11, respectively.\\ 
\indent The components of the flow velocity for the \textsf{hrez} model are illustrated by 2D colour maps in Fig. \ref{velx}. 
These snapshots of the flow were taken at 1,480$\,$s into the simulation, where the quasi-steady state has already 
developed. Each vertical 2D slice in Fig. \ref{velx} is taken 
at the same horizontal ($z$) position in the $x-y$ plane, at $z=0.94\times10^9\,$cm (i.\,e. in the middle of the 
domain, see Fig. \ref{comp-domain} for the domain geometry). The left, middle and right panels show the $x,y$ and $z$ components of the velocity, respectively. In the left 
panel, strong, buoyant up-flows are shown in shades of red, while cooler, dense down-drafts are shown in shades of blue. 

\par The convective boundaries are apparent in all the velocity components from the sudden drop in magnitude. The lower 
convective boundary is clearly distinguishable, but the upper boundary is more subtle with velocities above the boundary 
represented by slightly lighter shades of red and blue. In the middle and right panels, horizontal velocities are 
strongest near the convective boundaries (shown by extended patches of dark red and dark blue colours), this is 
indicative of the flow turning as it approaches the boundary. Gravity mode waves excited by turbulence in the convective 
region can be seen in the stable region above, and are shown by lighter shades of red and blue in the upper part of each panel.

\subsection{\corrnobf{Turbulent velocity spectrum}}\label{k53}

\corr{We investigate the degree to which our simulations are capturing the phenomenology of turbulence, including whether or not they have developed an inertial sub-range, by looking at velocity spectra of the modelled flows.} \corr{Spectra were calculated using a 2D fast Fourier transform\footnote{Using the Python package \textsc{numpy.fft.fft2}.} (FFT) of the vertical velocity in a horizontal plane at the mid-height of the convection zone. The results of this transform are presented in Fig. \ref{ke_spectra}, where the square of the transform, $\overline{\hat{V}^2}(k)$ is plotted as a function of the wave-number $k$. These spectra are time-averaged over several convective turnovers, and a 1D profile is obtained by binning the 2D transform within the $k_y-k_z$ plane, where $k_y$ and $k_z$ are the wave-numbers in the y and z directions, respectively ($k_y, k_z = 0, 2\pi,4\pi, \ldots , 2\pi (N/2)$, where $N$ is the number of grid points in one dimension, i.\,e. the resolution).}

\corr{A scaling of $(k^{5/3}/N)$ is applied to the velocity spectrum to compensate for its $k^{-5/3}$ dependence in the inertial range \citep{1941DoSSR..30..301K}. \corrdre{A plateau in the velocity spectra can be seen in all of the models. This plateau extends over the largest range in wave-numbers for the \textsf{vhrez} (cyan) case, $10 \lesssim k \lesssim 500$. Although this plateau in the spectrum is not a formal proof of the existence of an inertial range, it supports the fact that our simulations (at least in the \textsf{hrez} and \textsf{vhrez} cases) resolve appropriately the various ranges of the problem.}}

\corr{These velocity spectra thus demonstrate that our two highest resolution PPM simulations posses essential characteristics of a turbulent flow -- an integral scale, 
an inertial range obeying the $k^{-5/3}$ power law \corrdre{(at least for a sub-range of wave-numbers)}, and an effective Kolmogorov length-scale (represented by the grid scale). In our two lowest resolution runs, on the other hand, the plateau is either very short or not present, indicating that models with fewer than 512$^3$ zones are  probably not very accurate models of turbulence. This minimum desired resolution is in reasonable agreement with our estimate of the numerical Reynolds number in \S\ref{spatial_turb} ($\textrm{Re}_{\rm eff}\sim N_x^{\,4/3}$).}

\subsection{Mean Field Analysis of Kinetic Energy}\label{MFA}

\par A common method to study turbulent flows is to use the Reynolds-averaged Navier-Stokes (RANS)
equations. This reduction of multi-dimensional data into horizontal and time averaged one-dimensional (1D) radial profiles 
allows us to represent the data obtained from hydrodynamic simulations in the context of 1D 
stellar evolution models \citep{2014arXiv1401.5176M}. 

\begin{figure*}
\centering
\includegraphics[width=\textwidth]{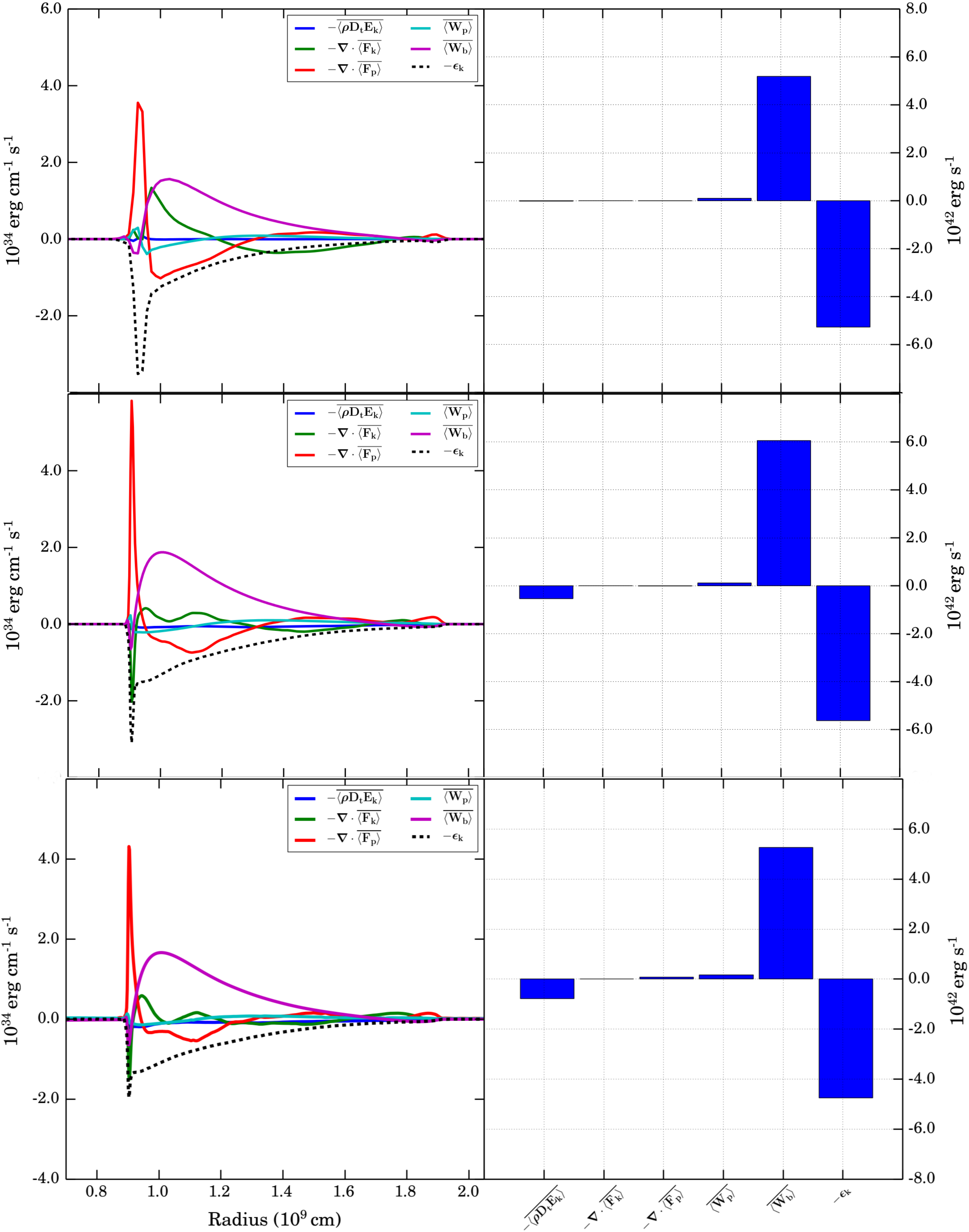}
\caption{\textsf{Left}: Decomposed terms of the mean kinetic energy equation (Eq. \ref{mmfka}), which have been 
horizontally averaged, normalised by the domain surface area and time averaged over the steady-state period. Time averaging 
windows are over 2,200$\,$s, 1,850$\,$s and 1,000$\,$s for the \textsf{lrez} (top), \textsf{hrez} (middle) and \textsf{vhrez} (bottom) models, 
respectively. \textsf{Right}: Bar charts representing the radial integration of the profiles in the 
left panel. \protect{This plot is analogous to the middle panels of fig. 8 in \citet{2013ApJ...769....1V}.}}
\label{tke_mfa}
\end{figure*} 

We use the RANS framework to calculate the terms of
the TKE equation (details given in Appendix \ref{app-tke}) and to analyse them.  
Momentum diffusion is not included in our simulations as we solve the inviscid Euler equations within the ILES paradigm. Instead, we infer TKE dissipation through the truncation errors that arise due to discretising these equations \citep{2007iles.book.....G}, this provides us with an effective numerical \corr{dissipation} ($\epsilon_k$ in Eq. \ref{mmfka}), which we compute from the residual energy in the TKE budget.

\subsubsection{Time-Averaged Properties of the TKE Budget}

\par The profiles of the mean TKE equation terms (Eq. \ref{mmfka}) for the
\textsf{lrez}, \textsf{hrez} and \textsf{vhrez} models are shown in the left panels of Fig. \ref{tke_mfa}, with the inferred viscous dissipation shown by a black dashed line. These profiles are time integrated over multiple convective turnovers and normalised by the surface area of 
the domain. Bar charts of the mean fields integrated over the domain are shown in the right panels.
Comparing the left panels of Fig. \ref{tke_mfa} to fig. 8 of \citet{2013ApJ...769....1V}, we see that the energetic properties 
of convection during carbon burning are very similar to oxygen burning.\\

\noindent {\em Time Evolution. --- } The Eulerian time derivative of the kinetic energy, $\rho \mathbf{D_t} E_k$, is small or negligible over the simulation  domain, implying that over the chosen time-scale the model is in a statistically steady state. \\

\noindent {\em Transport Terms. --- } The transport of kinetic energy throughout the convective region is determined by the two transport terms, the TKE flux, $\mathbf{F_k}$, and the acoustic flux, $\mathbf{F_p}$ \citep[see][for a detailed discussion on these terms]{2013ApJ...769....1V}.\\

\noindent {\em Source Terms. --- } Turbulence is driven by two kinetic energy source terms, $\mathbf{W_b}$ and $\mathbf{W_p}$. The rate of 
work due to buoyancy, $\mathbf{W_b}$ (density fluctuations), is the main source of kinetic energy within the convective region, while 
$\mathbf{W_p}$, the rate of work due to compression (pressure fluctuations or pressure dilatation) is small. 
In the convective zone, we generally have $\mathbf{W_b}>0$, as expected since it is the main driving term. Near the 
boundaries, however, there is a region where $\mathbf{W_b}<0$. These regions are where the flow decelerates (braking 
layer) as it approaches the boundary, as already found and discussed for oxygen burning in 
\citet{2007ApJ...667..448M} and \citet{2015ApJ...809...30A}. We note that the top braking layer is more 
extended than the bottom one. The top convective boundary width is also more extended. We come back to 
this point in \S\ref{thick}. \\

\noindent {\em Dissipation. --- } Kinetic energy driving is found to be closely balanced by viscous dissipation, $\epsilon_k$; a property consistent with the statistical steady state observed. 
 \corr{The time and horizontally averaged dissipation can be seen to extend roughly uniformly throughout the convective region but increases slowly in its amplitude with depth, tracking the RMS velocities. There is almost no dissipation in the stable layers, where velocity amplitudes are low and turbulence is absent. Finally, there are notable peaks in dissipation localised at the convective boundaries.  The dependence of these peaks on resolution is discussed next.}
 
\subsubsection{Resolution Dependence}
 \begin{figure*}
\includegraphics[width=0.49\textwidth]{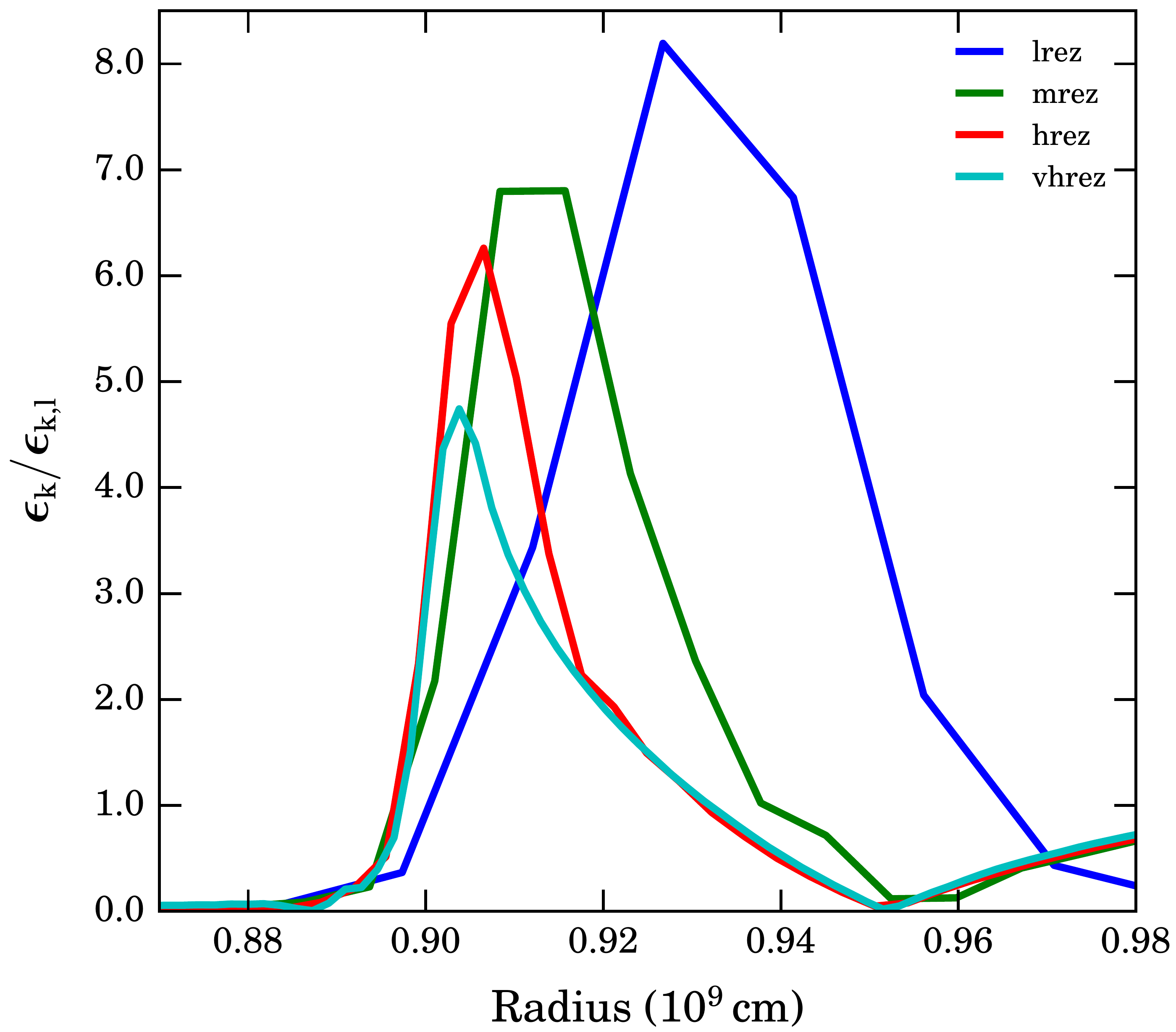}
\includegraphics[width=0.49\textwidth]{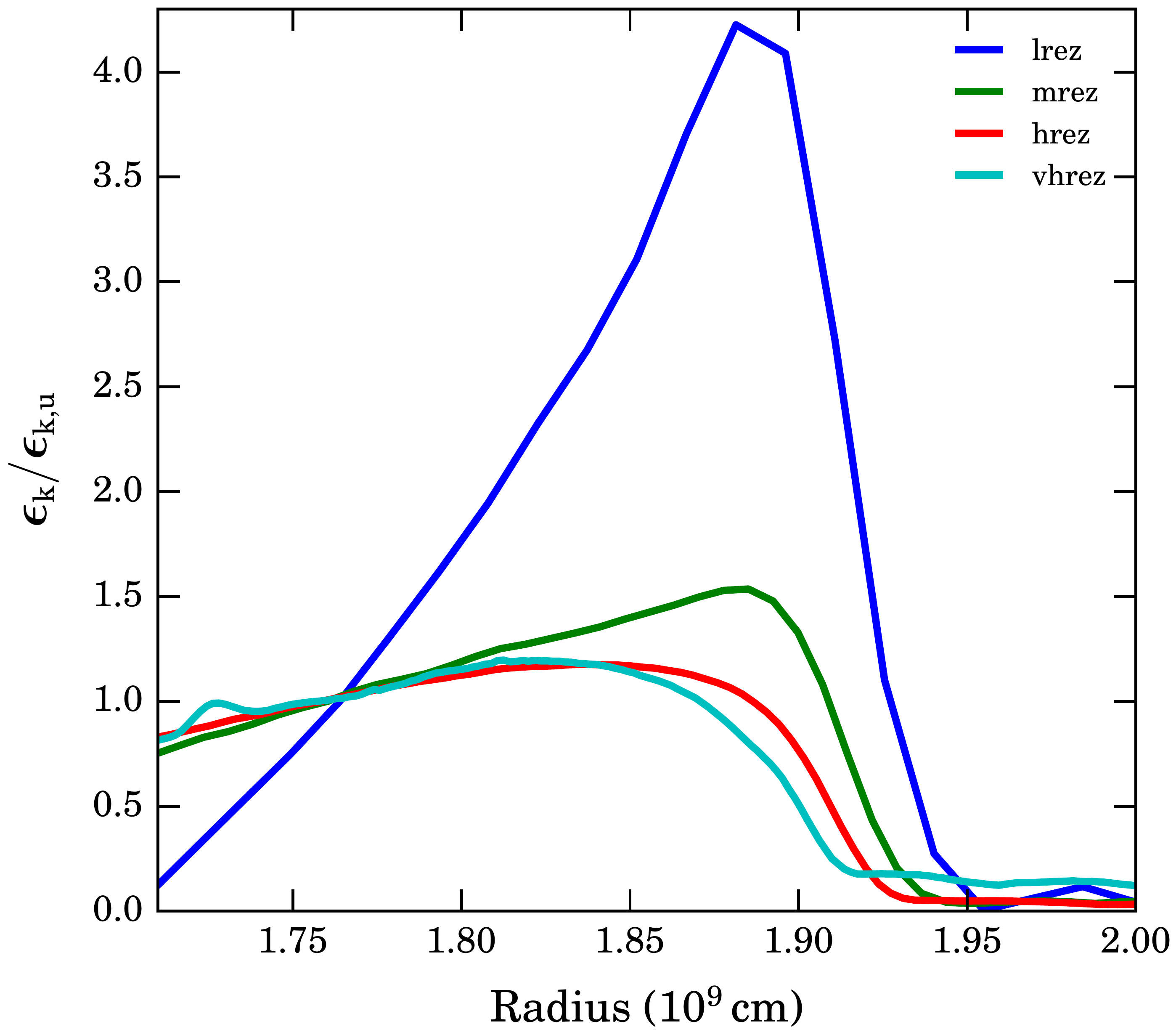}
\caption{Numerical dissipation inferred from the residual turbulent kinetic energy for the lower (left) and upper (right) convective boundary regions 
in the \textsf{lrez}, \textsf{mrez}, \textsf{hrez} and \textsf{vhrez} models. The dissipation at each boundary has been normalised 
by a value at a common position within the convective region near to the boundary. The 
\textsf{hrez} and \textsf{vhrez} residual profiles appear to be converging at the upper boundary, suggesting that the 
representative numerical dissipation here is physically relevant.}
\label{ek-boundary}
\end{figure*} 

We compare models of three different resolutions - the \textsf{lrez}, \textsf{hrez} and \textsf{vhrez} models, to 
determine if any of the physical results depend on
the chosen mesh size. Over the three resolutions, we find qualitatively similar results but there is significant 
deviation at the lower boundary region ($\sim0.9\times10^9\,$cm). 
\corr{A key question is whether or not our higher resolution models are able to capture the physics at boundaries accurately. }

At the lower convective boundary ($\sim0.9\times10^9\,$cm) a peak in dissipation appears at all resolutions (see dashed line in left panels of Fig. \ref{tke_mfa}). The peak decreases in amplitude and width with increasing resolution, indicating that the models are not converged numerically.

\par A comparison of the dissipation in this region for all resolutions is given in the left panel of Figure~\ref{ek-boundary}. Here the TKE dissipation is normalised by a value at a common position within the convective region 
near to the boundary. This highlights the relative decrease in this numerical peak with respect to a converged value in 
the convective region. A similar plot for the upper boundary is presented in the right panel of Figure~\ref{ek-boundary} shows that the 
dissipation at the boundary is smooth for both \textsf{hrez} and \textsf{vhrez} models. \corr{While in all cases the dissipation curves contain some variance due to the stochastic nature of the flow, the trend with resolution is clear.}

\subsection{Convective Boundary Mixing}\label{turb_entr}
\begin{figure*}
 \includegraphics[width=0.063\linewidth,height=0.42\linewidth]{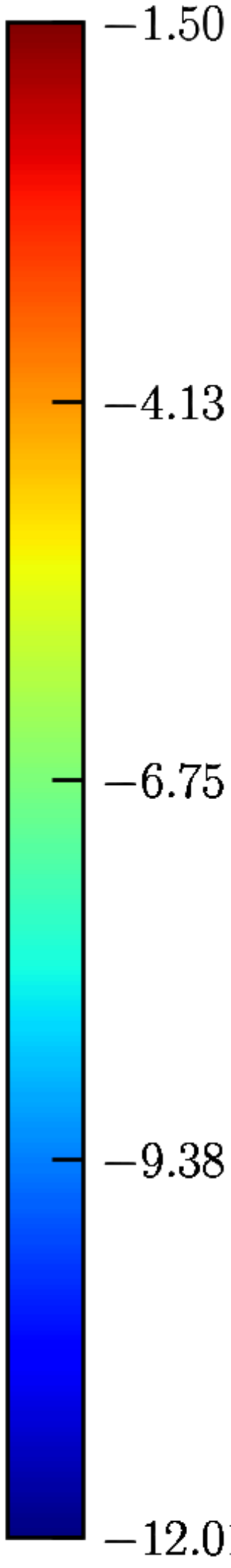}
 \includegraphics[width=0.42\linewidth]{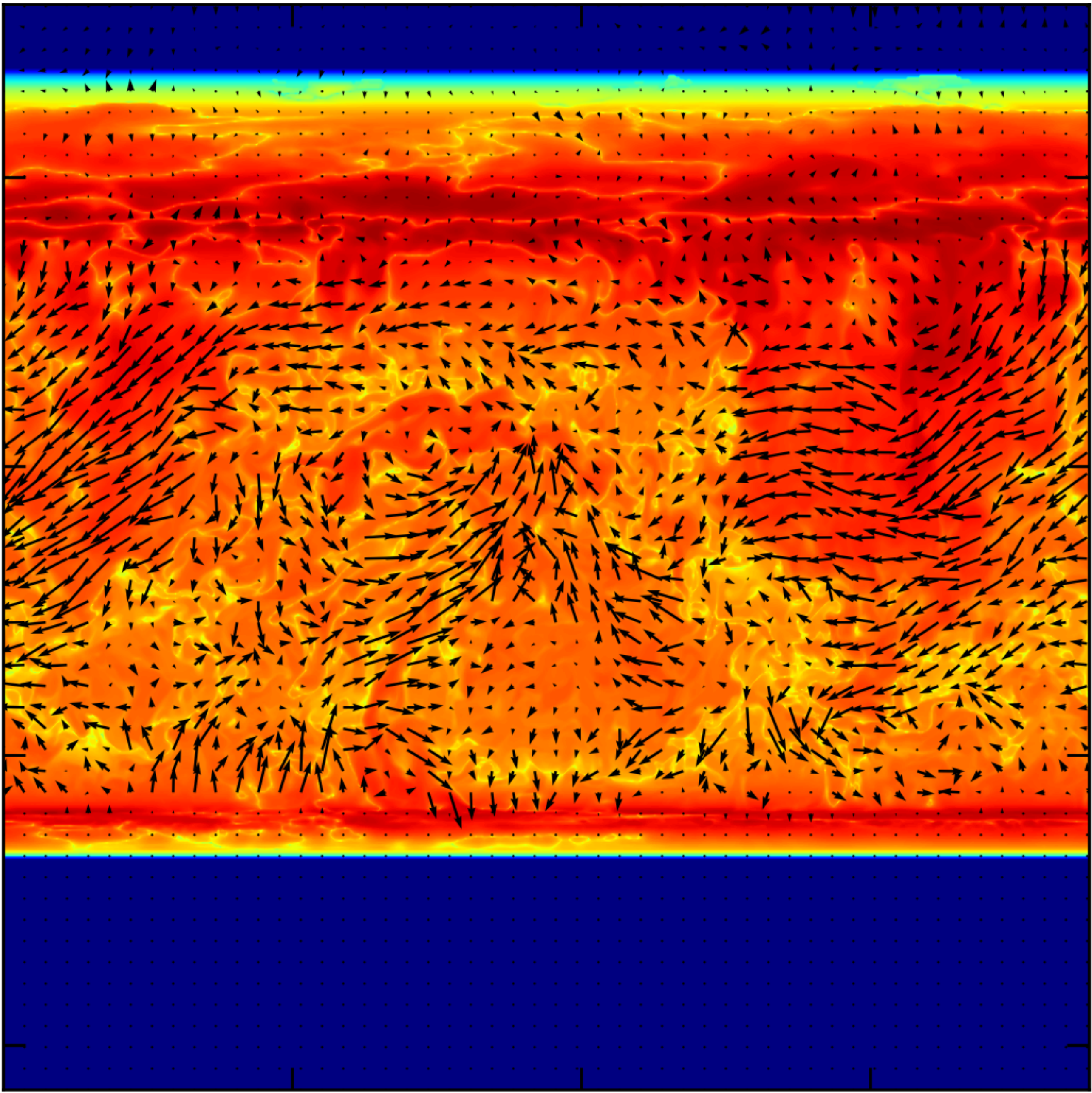} 
 \includegraphics[width=0.42\linewidth]{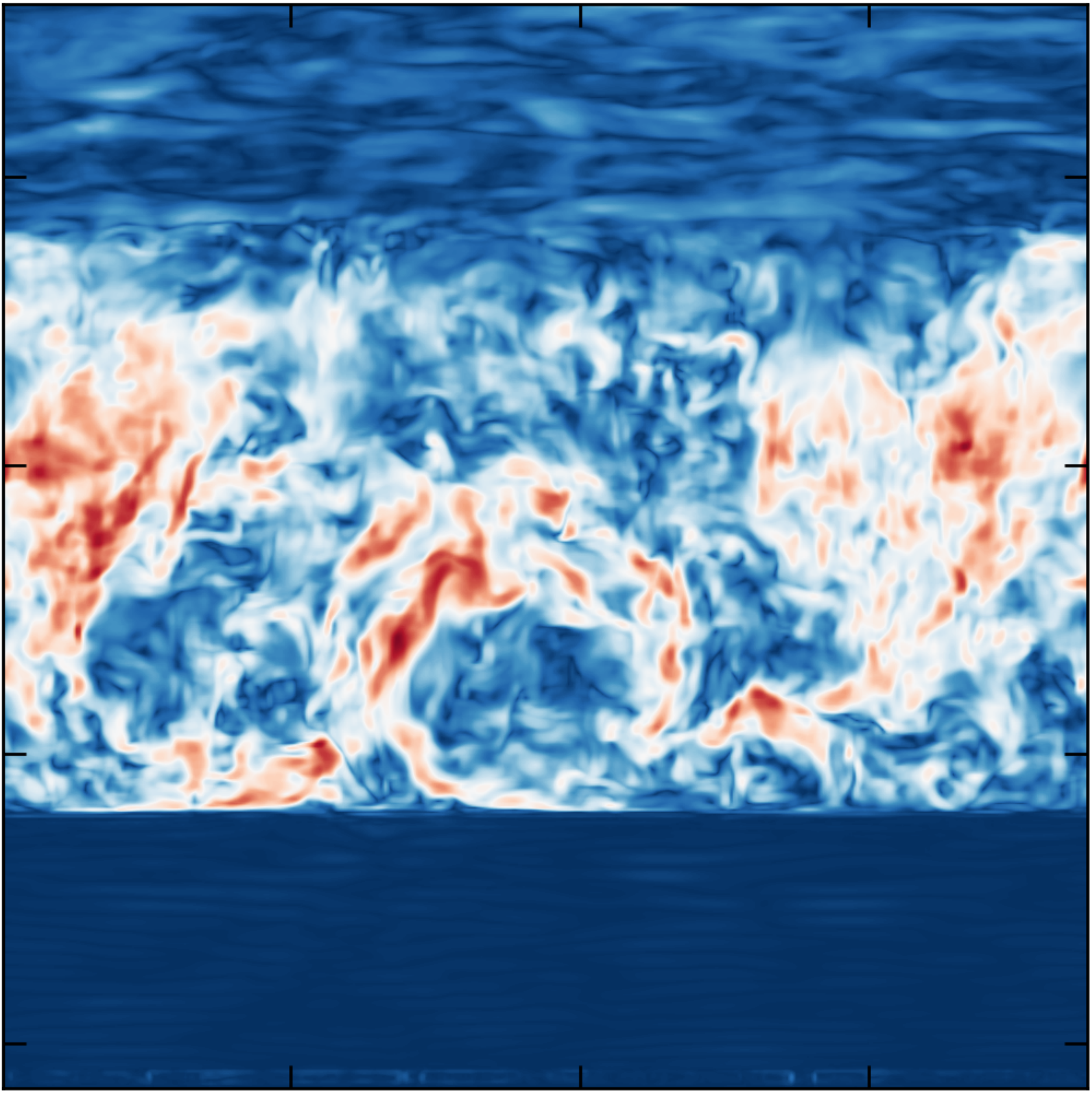} 
 \includegraphics[width=0.063\linewidth,height=0.42\linewidth]{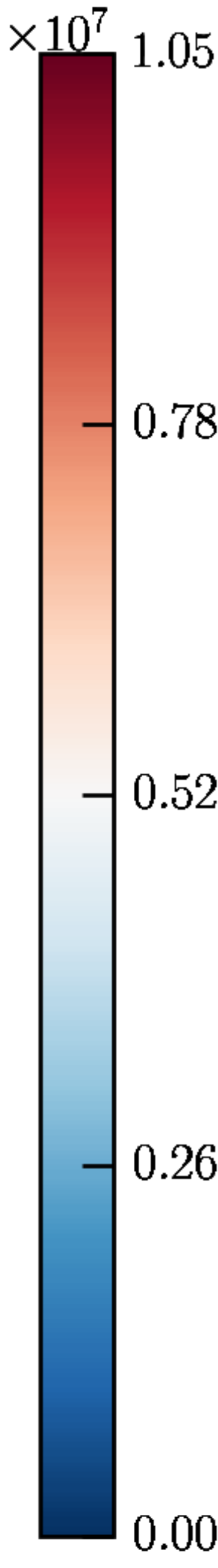}
\caption{\textsf{Left:} Vertical cross-section of the absolute average atomic weight fluctuations relative to their mean within 
the convective region. The colour map represents the logarithm of compositional fluctuations 
($|\overline{A}'/\overline{A}_0|$) \corr{relative to} the mean. Arrows show the vertical component and one horizontal 
component of the velocity vector field, $(v_x,v_y)$ \corr{(the vertical axis corresponds to the radial/vertical direction, see Fig.\,\ref{comp-domain})}. The direction of the arrows indicates the direction of 
this vector field in the x-y plane, and their 
length the magnitude of the velocity vector, $\sqrt{v_x^2 + v_y^2}$, at that grid point. \textsf{Right:} Vertical 
cross-section of the same velocity vector field plotted as 
 arrows in the left panel. \corr{The colour-map represents the velocity magnitude in cm\,s$^{-1}$.}
Both snapshots were taken at 2,820\,s into the \textsf{hrez} simulation. A movie of the velocity magnitude is available on this webpage: \textit{http://www.astro.keele.ac.uk/shyne/321D/convection-and-convective-boundary-mixing/visualisations/very-high-resolution-movie-of-the-c-shell/view}.}
\label{abar_cz}
\end{figure*} 

\par Entrainment events \citep[similar to entrainment events found for oxygen burning, see e.\,g. fig. 23 
in][]{2007ApJ...667..448M}  in the \textsf{hrez} model can be seen in the left panel of Fig. 
\ref{abar_cz} (see e.\,g. bottom left of convective zone where material from below the convective zone is entrained 
upwards or top corners of the convective zones where the material is entrained from the top stable layer). The left 
panel shows the average atomic weight fluctuations relative to their mean, with the velocity field in the 
$(x,y)$ plane over-plotted \corr{(the vertical axis corresponds to the radial/vertical direction, see Fig.\,\ref{comp-domain}). The right panel also shows 
the velocity magnitude ($\sqrt{v_x^2 + v_y^2}$)} for the same snapshot of the \textsf{hrez} model. In both panels, strong 
flows can be seen in the centre of the convective region and \corr{shear flows can be seen over the entire convective region. These shear flows have the greatest impact at the convective boundaries, where composition and entropy are mixed between the convective and radiative regions. Turbulent entrainment within the convective shell can also be inferred through the radial profile of the buoyancy work, whereby the positive work near the boundaries (e.g. the magenta curve of Fig. \ref{tke_mfa} at $\sim0.9\times10^9\,$cm) implies that that TKE of overturning fluid elements near the boundary does work against gravity to draw stable material into the convective region. This characteristic is explained in detail and seen in the buoyancy flux profiles of the oxygen burning shell in \citet{2007ApJ...667..448M} (see their \S7.2 and the top panel of fig. 25).} This is a very 
different picture from the parametrisations that are used to describe convective boundary mixing in  most modern 1D stellar evolution models. 

\corr{In this section, we start by estimating the position (and its time evolution) and thickness of the boundaries. We then interpret the time evolution of the boundary positions in the framework of the entrainment law. Finally, we compare the upper and lower boundaries.}

\subsubsection{Estimating Convective Boundary Locations}
\label{bound_loc}
\par Entrainment at both boundaries pushes the boundary position over time into the surrounding stable regions. 
In order to calculate the boundary entrainment velocities, first the convective boundary positions must be determined 
in the simulations. In the 3D simulations, the boundary is a two-dimensional surface and is not spherically symmetric as 
in 1D stellar models. 
In order to estimate the radial position of a convective boundary we first map out a two-dimensional horizontal boundary 
surface, $r_{j,k}=r(j,k)$, for $j=1,n_y$; $k=1,n_z$, where $n_y$ and $n_z$ are the number of grid points in the 
horizontal $y$ and $z$ directions. We estimate the radial position of the boundary at each horizontal coordinate to 
coincide with the position where the average atomic weight, $\bar{A}$, is equal to the average between the 
mean value of $\bar{A}$ in the convective and the corresponding radiative zones as defined in Eq.\,\ref{boundaryloc}.
\corr{The standard deviation of the position, $\sigma_r$, represents the amplitude of the fluctuations of the vertical position of the boundary across the horizontal plane due to the fact the the boundary is not a flat surface. }
\corr{Our method is a valid but not a unique way in which to calculate the boundary position. See \citet{1998JAtS...55.3042S,2004JAtS...61..281F,2007ApJ...667..448M,2011PhRvE..84a6311L,2011JAtS...68.2395S,2011JPhCS.318d2061V,2014AMS...1935.JGJG,2015JFM...778..721G} for a discussion of alternative definitions. The time evolution of the boundary position and its standard deviation are plotted in Fig.\,\ref{rad}.}

\begin{figure*}
\includegraphics[width=\textwidth]{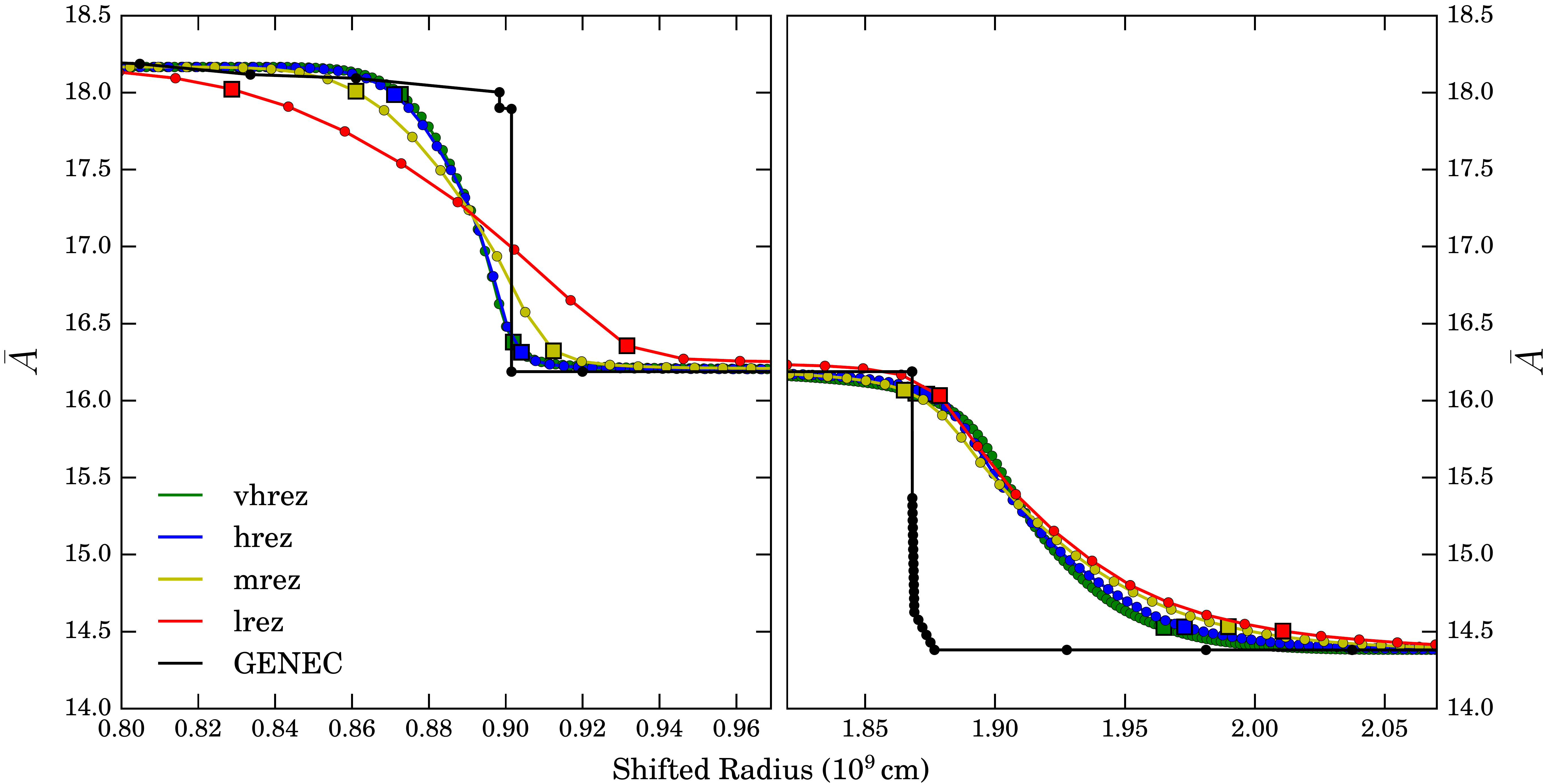}
\caption{Radial compositional profiles at the lower (\textsf{left}) and upper (\textsf{right}) convective boundary regions for the last time step of each model. 
\corr{The radius of each profile is shifted such that the boundary position, $\overline{r}$ (see \S\ref{bound_loc}), coincides with the boundary position of the \textsf{vhrez} model. 
In this sense, it is easier to assess the convergence of each model's representation of the boundary at the final time-step.} Individual 
mesh points are denoted by filled circles. Approximate boundary extent (width) is indicated by the distance between two filled squares for 
each resolution. The initial composition profile calculated using 
\textsc{genec} is shown in black (for a qualitative comparison only). See the corresponding text (\S\ref{thick}) for details of the definition of the boundary width.}
\label{figbw}
\end{figure*} 

\begin{figure*}
\centering
\includegraphics[width=\textwidth]{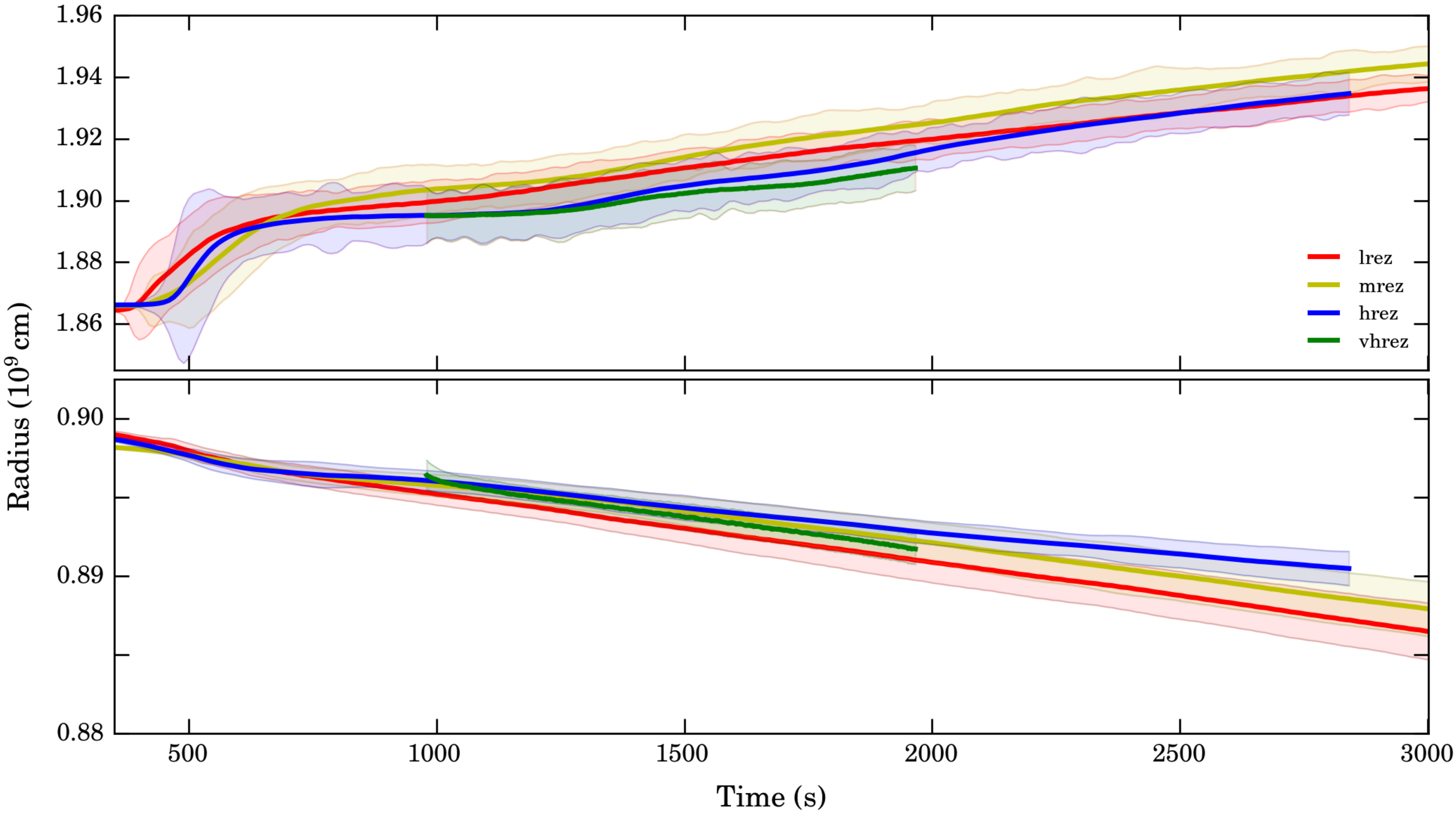}
\caption{Time evolution of the mean radial position of the convective boundaries, averaged over the 
horizontal plane for all four resolutions. Shaded envelopes are twice the standard 
deviation from the boundary mean location. \corr{The convective turnover time in these simulations is of the order of 1,000\,s}. \textsf{Top panel}: Upper convective boundary region. For increasing 
resolution, the average standard deviation, $\sigma_r$, in the estimated boundary position are the following percentages of the 
local pressure scale height: 3.3\%; 3.8\%; 4.3\%; and 4.5\%. \textsf{Bottom panel}: Lower 
convective boundary region. 
For increasing resolution the average standard deviation in the estimated boundary position are the 
following percentages of the local pressure scale height: 0.8\%; 0.7\%; 0.4\%; and 0.6\%. These shaded areas represent 
the variation in the boundary height due to the fact that the boundary is not a flat surface. This can be compared 
to the surface of the ocean not being flat due to the presence of waves.}
\label{rad}
\end{figure*} 

\subsubsection{Convective Boundary Structure}\label{sec:convective-boundary}\label{shap}

\par While stellar evolution codes describe a convective boundary as a  discontinuity (see the composition profile
in the right panel of Fig. \ref{1d3d}, for example), 3D hydrodynamic simulations show a more 
complex structure. A boundary layer structure is formed between the convective and stably stratified regions.
This can be seen from the apparent structure of the mean fields, 
at $\sim0.9\times10^9\,$cm and $\sim1.9\times10^9\,$cm, in the left panels of Fig. \ref{tke_mfa}, which represent the 
approximate locations of the lower and upper convective boundaries, respectively. 

\par The buoyancy in the convective boundary regions is negative, as seen in the $\mathbf{W_b}$  
profiles of Fig. \ref{tke_mfa}. 
In these regions, approaching fluid elements are decelerated and radial velocities greatly reduced.
As horizontal velocities increase, the plumes turn around and fall back into the convective region.
This is similar to the description by \citet{2015ApJ...809...30A} (see their fig. 5 and text therein).
\vspace{-0.275cm}
\subsubsection{Convective Boundary Thickness Estimates}\label{thick}

\begin{figure*}
\centering
\includegraphics[width=0.75\linewidth]{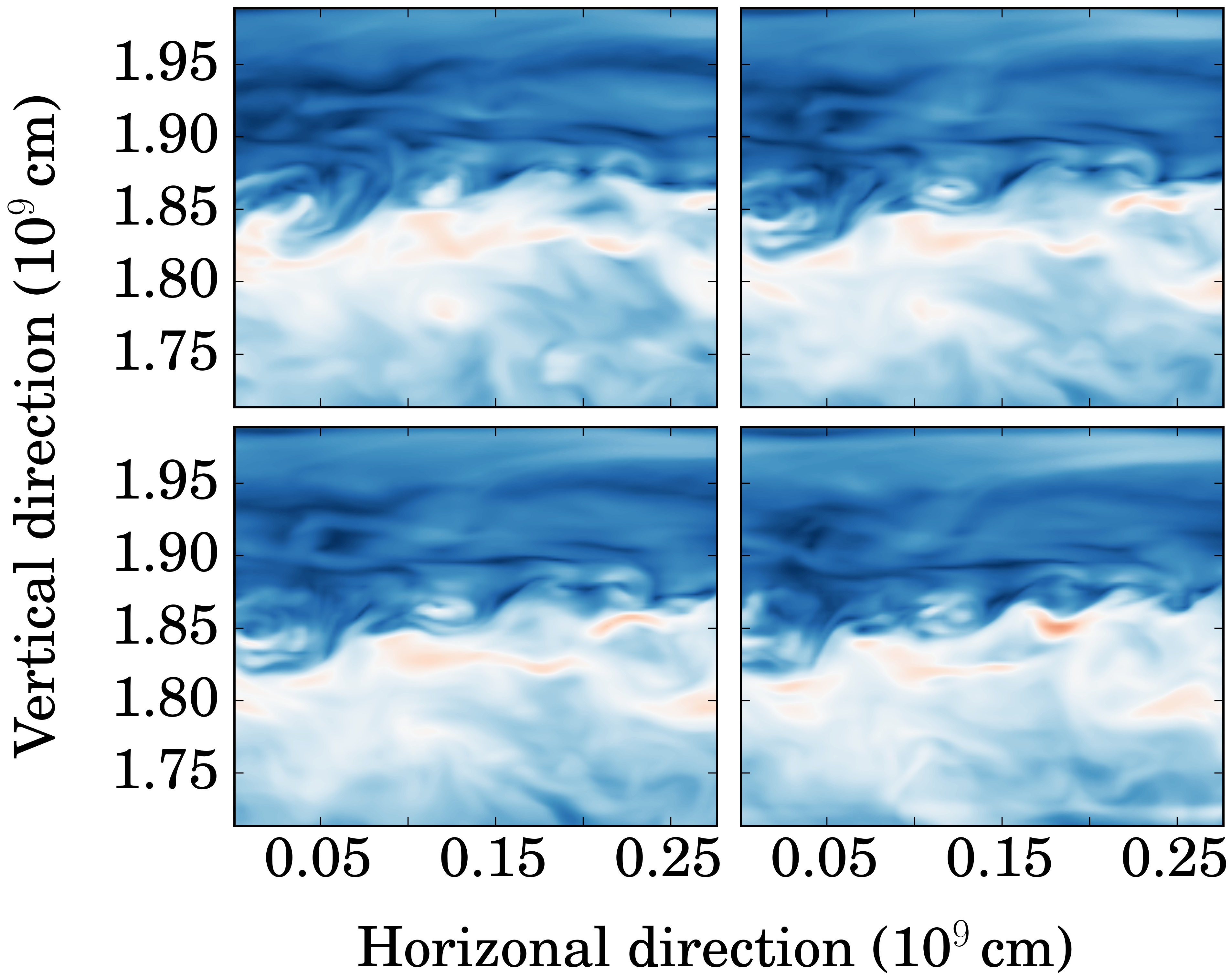}
\includegraphics[width=0.1\linewidth]{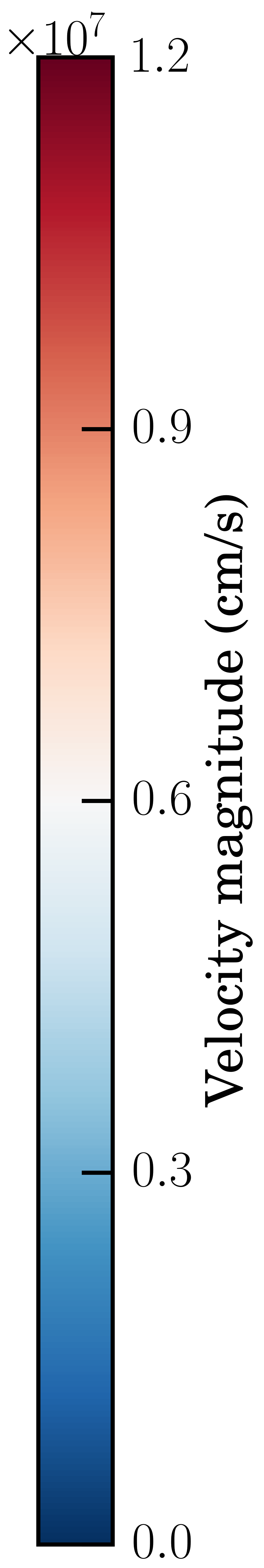}
\caption{Sequential vertical cross-sections in the x-y plane of the velocity magnitude, $\sqrt{v_x^2+v_y^2+v_z^2}$, across the left section of the upper convective boundary for the \textsf{vhrez} simulation. Snapshots are taken at 1565\,s (upper left), 1570\,s (upper right), 1575\,s (lower left) and 1580\,s (lower right). The colour bar presents the values of the velocity magnitude in units of cm\,s$^{-1}$. Each panel reveals \corrdre{shear mixing occurring across the boundary interface. The Kelvin-Helmholtz instability is a promising candidate for generating this type of mixing.}}
\label{kh_zoom}
\end{figure*} 

\par We estimate the thickness of the convective boundaries using the jump in composition, $\bar{A}$, between convective and stable regions. 
We denote the average composition (averaging removes stochastic fluctuations in composition) in the, lower stable, convective and upper stable regions as, $\bar{A}_l$, $\bar{A}_c$ and $\bar{A}_u$, respectively.
We consider the boundary region to extend between 99\% and 101\% of the respective positions coincident with such compositional values. For each boundary,
we signify such values by the appendage of a subscript $-$ (99\%) or $+$ (101\%) to the composition of each region.
Explicitly, the lower boundary thickness is defined as,

\begin{equation}\label{deltarl}
\delta r_l=r\left(\,\bar{A}_{c_{+}}\right)-r\left(\,\bar{A}_{l_{-}}\right).
\end{equation}

The upper boundary thickness is similarly defined as,

\begin{equation}\label{deltaru}
\delta r_u=r\left(\,\bar{A}_{u_{+}}\right)-r\left(\,\bar{A}_{c_{-}}\right).
\end{equation}

In addition, we also considered defining the boundary thickness using gradients in composition and entropy, and the jump in entropy at the boundary.
We found that these other methods gave quantitatively similar results.
In Fig. \ref{figbw}, we illustrate the estimation of the boundary thickness using Eqs. \ref{deltarl} and 
\ref{deltaru} for the final time-step of each simulation. \corr{The radius of each profile has been shifted, such that the boundary position, $\overline{r}$ (see \S\ref{bound_loc}),
of each model coincides with the boundary position of the \textsf{vhrez} model. With such a shift, it is easier to assess the dependence of the boundary shape on resolution.}

\par The extents of the convective boundaries are marked by filled squares for each simulation. Filled circles represent the individual mesh points, indicating the resolution of each simulation. Note that, the composition profile labelled as model \textsf{GENEC} is from the 1D stellar model, and was used as part of the initial conditions for all of the 3D models, so serves only as a qualitative comparison. The exact thickness of each boundary is shown in Table \ref{tabcbm}, along with their fraction of the local pressure scale height.

\begin{table}
\caption{Table summarising bulk and boundary region properties for each model. $v_{rms}$ - global RMS convective velocity (cm\,s$^{-1}$); $\ell_c$ - convective region height (cm); $v_e$ - entrainment 
velocity (cm\,s$^{-1}$); $\delta r$ - boundary region width (cm); $\delta r/H_p$ - ratio of the boundary region width to the average pressure scale height across the boundary; $\tau_b$ - boundary entrainment time (s); $\textrm{Ri}_{\rm B}$ - bulk Richardson number. Values in brackets correspond to the lower boundary.}
\hskip-2mm\begin{tabulary}{\textwidth}{c || c c c c}
\hline \hline\\
 & \textsf{lrez}  &  \textsf{mrez}  &  \textsf{hrez}  &  \textsf{vhrez} \\\\
\hline \hline\\
\textsf{$v_{rms}$} & 3.76 & 4.36 & 4.34 & 3.93 \\
\scriptsize \textsf{$(10^6)$}&&&&\\\\
\textsf{$\ell_{c}$} & 1.08 & 1.04 & 1.03 & 1.09 \\
\scriptsize \textsf{$(10^9)$}&&&&\\\\
\textsf{$v_{e}$} & 1.78 (-0.44) & 2.01 (-0.39) & 2.15 (-0.30) & 1.59 (-0.46) \\
\scriptsize \textsf{$(10^4)$}&&&&\\\\
\textsf{$\delta r$} & 13.2 (10.3) & 12.5 (5.1) & 9.9 (3.3) & 9.6 (2.9) \\
\scriptsize \textsf{$(10^7)$}&&&&\\\\
\textsf{$\delta r/H_p$} & 0.41 (0.36) & 0.36 (0.17) & 0.29 (0.11) & 0.28 (0.10) \\\\\\
\textsf{$\tau_b$} & 7.4 (23.4) & 6.2 (13.1) & 4.6 (11.0) & 6.0 (6.3) \\
\scriptsize \textsf{$(10^3)$}&&&&\\\\
\textsf{$\textrm{Ri}_\textrm{B}$} & 29 (370) & 21 (259) & 20 (251) & 23 (299) \\\\
\hline \hline
\end{tabulary}
\label{tabcbm}
\end{table}

In Fig. \ref{figbw} \corr{(right panel)}, it can be seen that the composition gradient at the top boundary is nearly converged between all resolutions and varies only mildly between the lowest resolution case and the other models. 
The composition gradient at the lower boundary \corr{(left panel)}, on the other hand, varies significantly between the \textsf{lrez} and \textsf{hrez} models, while between the \textsf{hrez} and \textsf{vhrez} the boundary shape appears to have nearly converged although is still narrowing slightly. These  trends are confirmed by the quantitative estimates of the boundary widths presented in Table \ref{tabcbm}.\\ 
\indent \corr{The thickness determined from the abundance gradients
is larger than the standard deviation, $\sigma_r$, of the 
boundary location} (corresponding to the mid-points of the abundance gradients plotted in Fig. \ref{figbw}) shown as 
shaded areas in Fig. \ref{rad}. This is expected since the fluctuations of the boundary location do not take into 
account its thickness or width, but only the location of its centre (mid-point). These fluctuations of the boundary location 
can be compared to fluctuations in the height of the ocean surface due to the presence of waves. The fact 
that the width determined from the abundance gradients (given in Table \ref{tabcbm}) is significantly larger means that 
there is mixing across the boundary. \corrdre{A promising candidate for this type of mixing is the Kelvin-Helmholtz instability which would give rise to the shear motions seen in Fig. \ref{kh_zoom}. This figure shows sequential slices of the flow velocity across the left section of the upper convective boundary region. Such shear mixing is induced by 
plumes rising from the bottom of the convective region and turning around at the boundary \citep[see also the shear layer in fig. 5 of][]{2015ApJ...809...30A}.} Mixing also occurs through plume impingement or penetration with the boundary. Some mixing may also occur through the presence of gravity waves which propagate through the stable region. It is not expected that the upper boundary gradient will steepen, as this would support more violent surface waves whose non-linear dissipation would tend to broaden the gradient, resulting in a negative feedback loop between these two processes.\\
\indent It is important to remember that the boundary widths given in Table \ref{tabcbm} are only estimates. The key finding are (1) that  the lower boundary has a narrower width compared to the upper, and (2) the widths are relatively well converged between the {\sf hrez} and {\sf vhrez} models.

\subsubsection{Convective Boundary Evolution and Entrainment Velocities}

\par The variation in time of the average surface position, $\overline{r}$, of both boundaries is shown for all 
models in Fig. \ref{rad}. Positions are shown as solid lines and twice the standard deviation as the surrounding shaded 
envelopes. Following the initial transient 
(\textgreater$\,1,000\,\textrm{s}$) a quasi-steady expansion of the convective shell proceeds. 
We obtained the entrainment velocities, $v_e$, given in Table \ref{tabcbm} using a linear fit to the time evolution of the 
boundary positions \corr{during the quasi-steady phase}. These velocities are very high. If one multiplies 
them by the life-time of carbon shell burning (of the order of 10 years), the convective 
boundaries would move by more than $10^{10}$\,cm, which would lead to dramatic consequences for the 
evolution of the star.  Note, though, that the driving luminosity of the shell was boosted by a factor of 1,000. We 
will come back to this point in \S\ref{comp1d3d}.
  
\subsubsection{The Equilibrium Entrainment Regime}\label{equil}

\par In the equilibrium entrainment regime \citep{2004JAtS...61..281F,2014AMS...1935.JGJG}, the time-scale
for the boundary migration, $\tau_b$, is comparable to or larger than the turbulent transit time-scale, $\tau_c$ (\S\ref{sec:init-conditions}). Therefore, in this regime, the entrainment process is sampling the entire spectrum of turbulent motions over the inertial range rather than being sensitive to individual
turbulent elements, such as in strong, individual outliers events. This simplifies the development of mixing models within this regime.
The boundary entrainment velocity $v_e = d\,\overline{r}/dt$ is defined in terms of the mean boundary position $\overline{r}(t)$. 
We define the boundary mixing time-scale as $\tau_b = \delta r/|v_e|$, where $\delta r$ is the boundary thickness (Table \ref{tabcbm}), which we define in \S\ref{thick}. We find $\tau_b/\tau_c$ ratios for the upper convective boundary of $13.4,\, 13.1,\, 9.8$ and $11.7$ for the \textsf{lrez}, \textsf{mrez}, \textsf{hrez} and \textsf{vhrez} models, respectively, \corr{placing all of these boundaries firmly in the equilibrium regime.}

\subsubsection{The Entrainment Law}\label{sec:entr_law}

The time rate of change of the boundary position due to turbulent entrainment (the 
entrainment velocity), $v_e$, \corr{has been found to scale as a power of the bulk Richardson number for a wide range of conditions} \citep[e.\,g.][]{2014AMS...1935.JGJG}. \corr{This relationship is often referred to as an {\em entrainment law} in the meteorological and atmospheric 
and is typically written as}:

\begin{figure}
\includegraphics[width=0.5\textwidth]{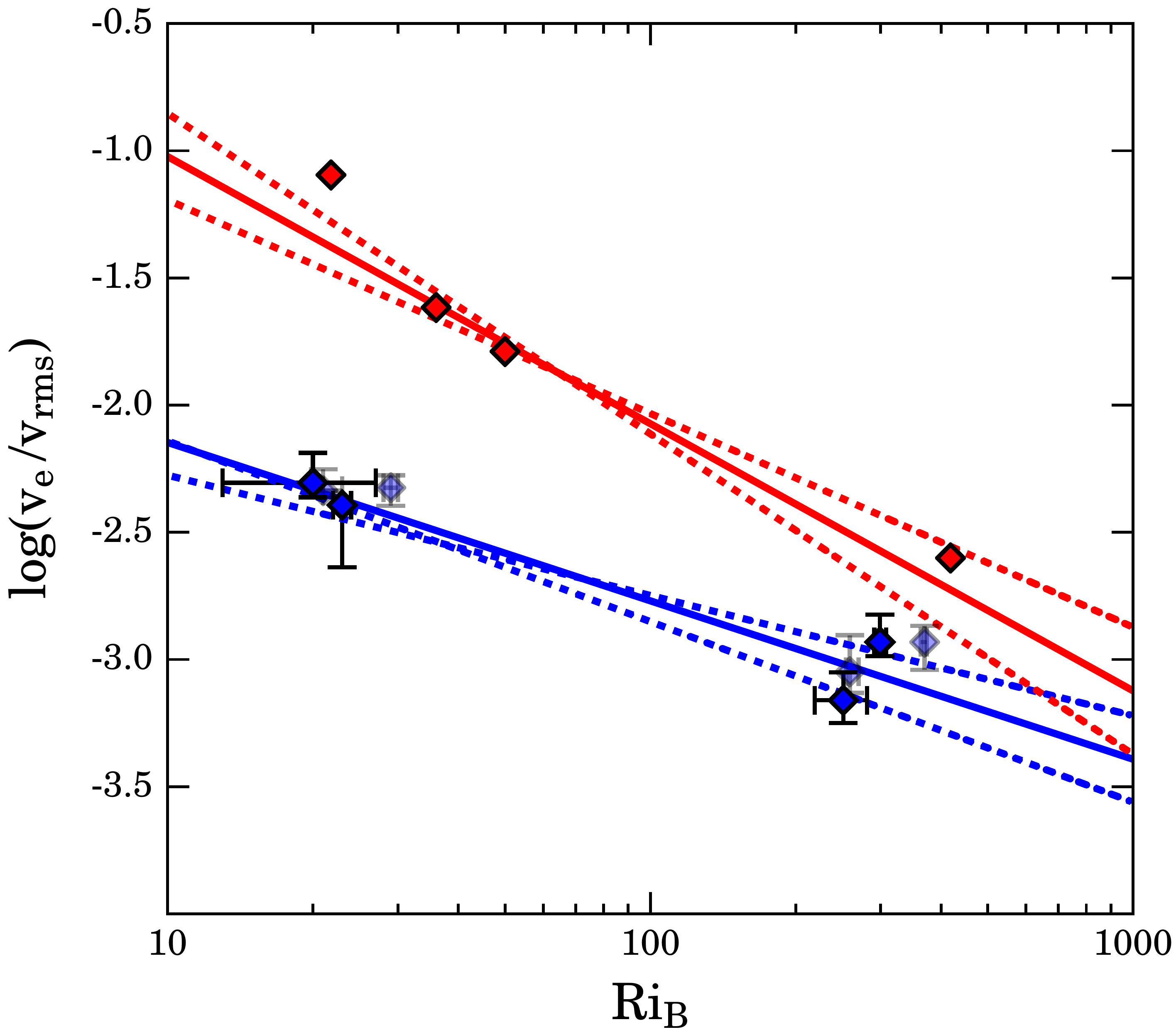}
\caption{Logarithm of the entrainment speed \corr{for high P\'eclet number simulations,} normalised by the RMS turbulent velocity versus the bulk Richardson 
number. Red points represent data obtained in the study by \citet{2007ApJ...667..448M} and blue 
points represent data obtained in this study, \corr{transparent points represent the values for the \textsf{lrez} and \textsf{mrez} models, which are not included in best fit power law shown by the blue solid line. The blue dashed lines show the best fit to the extremes of the error bars of the \textsf{hrez} and \textsf{vhrez} models. The red solid line is the best fit power law from a linear regression of the oxygen shell data and the red dashed lines show the error in the computed slope.}}
\label{entrain}
\end{figure} 

\begin{equation}\label{entr_law}
\frac{v_e}{v_{rms}}=A\, \textrm{Ri}_{\rm B}^{-n}.
\end{equation}

\par Many LES (e.\,g. \citealt{1980BoLMe..18..495D}) and laboratory 
(e.\,g. \citealt{2010NPGeo..17..187C}) studies have found similar values for the coefficient, $A$, 
typically between 0.2 and 0.25, although experimental measures have been more uncertain.

\par The exponent, $n$, is often found to be close to unity for convectively  driven  turbulence \citep[e.\,g.][]{1991AnRFM..23..455F,2001BAMS...82..283S}, a result that follows from basic energetic considerations. \corr{On the other hand, in a recent DNS study, \citet{2013JFM...732..150J} showed that $A\approx 0.35$ and $n=1/2$ for shear-driven entrainment.} 

\par We compare  the bulk Richardson number (Eq. \ref{rib}) of our 3D simulations to the initial conditions from the 1D stellar model \citep{2016PhyS...91c4006C} and 3D oxygen burning simulations from  \citet{2007ApJ...667..448M}.
From the 1D 15$\,\textrm{M}_\odot$ stellar model of \citet{2016PhyS...91c4006C}, used as initial conditions in these simulations, 
the bulk Richardson numbers of the carbon burning shell are $\textrm{Ri}_{\rm B}^{\,u}\sim1,440$ and 
$\textrm{Ri}_{\rm B}^{\,l}\sim2.0\times10^4$ at the upper and lower convective boundaries, respectively. 
While, for our 3D \textsf{vhrez} model (see Table \ref{tabcbm}), we found $\textrm{Ri}_{\rm B}^{\,u}\sim23$ and 
$\textrm{Ri}_{\rm B}^{\,l}\sim299$. The lower values we obtain in 3D are mainly due to the fact that we 
boosted the luminosity by a factor of 1,000. This is further discussed in \S\ref{comp1d3d}.

\par \corr{The entrainment speed (normalised by the RMS velocity) is plotted as a function of the bulk Richardson number in Fig. \ref{entrain}. Red points represent the data obtained in the study by \citet{2007ApJ...667..448M}, the solid red line is a best fit power law to the data following a linear regression, and the red dashed lines show the error in the computed slope. Blue opaque points represent the values obtained in the \textsf{hrez} and \textsf{vhrez} models and blue transparent points are the values obtained in the \textsf{lrez} and \textsf{mrez} models. We obtain a best fit power law to the \textsf{hrez} and \textsf{vhrez} data and the extremes of their error bars, shown by the solid blue line and dashed blue lines, respectively. The corresponding intercept and slope of this best fit denote the entrainment coefficient, $A = 0.03\,(\pm0.01)$, and entrainment exponent, $n = 0.62\,(+0.09/-0.15)$, respectively. 
The value we obtain for the entrainment exponent, $n$, falls between
the two scaling relations ($1/2\leq n \leq 1$).
Our value for the coefficient, $A$, however, differs  from all of the values found in the literature. A larger dataset is desired with which to explore in more detail the parameter space. Interestingly, the bulk Richardson numbers are similar between the carbon and oxygen shell models, and in particular, the lower convective boundaries both have higher values than the corresponding upper boundaries.
The difference in the best fit values of $A$ indicates that the oxygen shell is more efficient in converting kinetic energy into mixing. The difference in the best fit values of $n$ indicates that there may be a second parameter besides Ri${\rm_B}$ that is varying between the top and bottom of the convective shells, and in different ways, between the carbon and oxygen shell models.  Finally, it must be reiterated that the ambiguity associated with calculating Ri${\rm_B}$ is likely to account for some of the discrepancy.}\\

\vspace{-0.8cm}
\subsubsection{Comparison of Upper and Lower Convective Boundary Properties}

\par Summarising the boundary properties determined so far for the \textsf{hrez} model (see Table \ref{tabcbm}),
the upper boundary region has a typical width of 
$9.9\times10^7\,\textrm{cm}$, entrainment speed of 
$2.2\times10^4\,\textrm{cm}\,\textrm{s}^{-1}$ and bulk Richardson number of 20. The 
lower boundary region typically has a   
width of $3.3\times10^7\,\textrm{cm}$, entrainment speed of $3\times10^3\,\textrm{cm}\,\textrm{s}^{-1}$  and bulk 
Richardson number of 251. 
We thus have a consistent picture of the lower boundary being narrower, having a slower entrainment velocity and being stiffer (higher Ri$_{\rm B}$) 
compared to the upper boundary by a factor of about 3, 7 and 13 respectively.

\subsection{Comparing Convective Boundary Mixing Between 1D and 3D Models}\label{comp1d3d}

Upon comparing our results to the 1D GENEC stellar evolution models, we find that our boundary widths are much larger and the 
boundary structures are very different from those calculated using strict Ledoux or Schwarzschild boundaries. The results of this and
similar 3D hydrodynamic studies \citep{2007ApJ...667..448M, 2013ApJ...769....1V, 2015ApJ...798...49W, 
2015ApJ...809...30A,  2015ApJ...808L..21C, 2017MNRAS.465.2991J} call for improved convective boundary mixing prescriptions 
in 1D stellar evolution models. 

\par An approximate relation can be obtained between Ri$_{\rm B}$ and the luminosity, allowing the determination
of convective boundary stiffness in 1D stellar models. Considering Eq. \ref{entr_law} \corr{(with $n=1$)}, the relation $v_{\rm rms} \propto L^{1/3}$ (assuming $\epsilon_{\rm nuc}\sim v_{\rm rms}^3/\ell$; \citealt{1941DoSSR..30..301K}, and $L=\int\epsilon_{\rm nuc}\,\textrm{d}m$) and that the \corr{entrainment rate scales roughly linearly 
with the luminosity of the shell over the quasi-steady state \citep{2017MNRAS.465.2991J}, which implies that $v_e\propto L$,} we obtain that Ri$_{\rm B} \propto L^{-2/3}$. Interestingly, we find the 
same dependence when starting from the formula for Ri$_{\rm B}$ (given in Eq. \ref{rib}), considering that the buoyancy jump 
remains constant (which is reasonable for a given initial stratification) and that $v_{\rm rms} \propto L^{1/3}$.
Using the relation 
Ri$_{\rm B} \propto L^{-2/3}$, the boost of a factor of 1,000 in the luminosity of our 3D models thus implies a reduction by a factor of 100 in Ri$_{\rm B}$. 
This brings the values of the bulk Richardson number between our 3D and 1D models of the carbon shell into a reasonable agreement (see \S\ref{sec:entr_law}).
\corr{A complication involves calculating the buoyancy jump needed for the bulk Richardson number since it is not precisely defined in a complex, stratified situation like a stellar interior -- the length-scale used for this integration is therefore somewhat arbitrary.} 

\par Another important point is that we confirm with the 3D simulations that the 
lower boundary is stiffer than the upper boundary, by a factor of about 13 \corr{in terms of Ri$_{\rm B}$}.
The fact that the 
entrainment velocity at the lower boundary is a factor of about 7 smaller than at the upper 
boundary is partly explained by the fact that the horizontal velocities at the lower boundary are 
higher than at the upper boundary (see Fig. \ref{urms-fluct}). 

\par  Theoretical relations like the entrainment law will be 
needed to determine entrainment velocities for different 
burning stages and their various phases. This can be achieved by first estimating the bulk Richardson number of a given convective
boundary from the luminosity, as described above. Then, one can approximate a turbulent RMS velocity using 
the velocity calculated from MLT or a similar method. Finally, Eq. \ref{entr_law} can be used with suitable values for the 
entrainment coefficient and exponent to estimate the entrainment velocity of the mentioned convective boundary (e.\,g. 
\citealt{1980BoLMe..18..495D, 1991AnRFM..23..455F, 
2001BAMS...82..283S, 2007ApJ...667..448M, 2010NPGeo..17..187C}). 

\section{Comparison to Other Simulations}\label{other_sims}

As found by \citet{2016PhyS...91c4006C}
using the same 1D stellar model used for these hydrodynamic simulations, the lower convective boundary is stiffer 
than the upper boundary as 
determined by the bulk Richardson number. Our higher resolution 3D models produce comparable results for the bulk Richardson number and for the \textsf{hrez} model we obtain 
values of 20 and 251 for the upper and lower boundaries, respectively.\\
\indent \citet{2007ApJ...667..448M} simulated the oxygen shell of a 23$\,\textrm{M}_\odot$ star in spherical coordinates, also using the \textsc{prompi} code. The driving of the carbon
shell we simulate is similar to their oxygen shell, owing to the fact that we boosted the luminosity. We find that the 
profiles of the velocity components
are comparable between the two models. As shown by our Fig. \ref{entrain} and their fig. 26, we find similar estimates 
of the bulk Richardson numbers, \corr{while the values of 
the constants $A$ and $n$ (from Eq. \ref{entr_law}) differ, this is somewhat expected as the oxygen shell engulfs the neon burning shell and complex multiple shell burning proceeds.}\\
\indent We obtain a turbulent kinetic energy budget that is in agreement with that of spherical simulations of the oxygen shell in a 23$\,\textrm{M}_\odot$ star by \citet{2013ApJ...769....1V}. In such an energy budget we see a statistically steady state of turbulence over four convective turnovers. Predominantly, this 
is driven from the bottom of the shell by 
a positive rate of work due to buoyancy and dissipated at the grid scale by a numerical viscosity.\\
\indent In recent full $4\pi$ simulations of the oxygen burning shell in a 25$\,\textrm{M}_\odot$ star, 
\citet{2017MNRAS.465.2991J} find a 2$\sigma$ fluctuation in their calculation of 
the convective boundary of 17\% of the local pressure scale height. This is larger than the 
horizontal fluctuation in our estimation of the upper boundary of the carbon shell; a 2$\sigma$ fluctuation of 4.3\%
of the local pressure scale height (Fig. \ref{rad}). This difference could be due to the maximum 
tangential velocity gradient method that \citet{2017MNRAS.465.2991J} use to estimate the boundary positions, 
which differs from the method described in \S\ref{bound_loc}. We find comparable 
magnitudes of the velocity components (see our Fig. \ref{urms-fluct}, and their fig. 11), and also 
similar Mach numbers for the flow (see our Table \ref{tab1}, and their table 1). This could be 
in part due to the fact that our boosted energy generation rate ($\sim3\times10^{12}\,$erg$\,$g$^{-1}\,$s$^{-1}$) 
is comparable to the rate used in their \textsc{PPMstar} \citep{2015ApJ...798...49W} simulations. The relative magnitude of the 
radial velocity component in Fig. \ref{urms-fluct} is higher than that of \citet{2017MNRAS.465.2991J},
and our horizontal velocity does not possess the same symmetry as their tangential velocity. The latter could be due 
to the difference in geometries between the two simulations. \citet{2017MNRAS.465.2991J}
also observe entrainment at the upper convective boundary of their oxygen shell. Their velocity of the upper
boundary due to entrainment is lower than the entrainment velocity we estimate by over an order of magnitude.
One reason for this difference could be that the oxygen shell boundary is much stiffer than the carbon
shell boundary, due to a smaller jump in buoyancy over the boundary (RMS turbulent velocities are similar). We determined this difference in boundary stability through the difference in the peak squared Brunt-V\"{a}is\"{a}l\"{a}
frequencies. The value for the upper boundary of the carbon shell ($\sim 0.05\,$rad$\,$s$^{-2}$) is an order of 
magnitude smaller than 
that of the oxygen shell. This could explain the order of magnitude difference in entrainment velocity assuming that
the oxygen shell simulations also follow an entrainment law of the form of Eq. \ref{entr_law}.

\section{Conclusions}\label{conc}

3D hydrodynamic simulations that represent the second carbon shell of a 15$\,\textrm{M}_\odot$ star have been performed, using the \textsc{prompi} code. The initial conditions used were finely mapped profiles of the carbon shell structure from a  15$\,\textrm{M}_\odot$, solar metallicity, \corr{non-rotating}   stellar model calculated by \citet{2016PhyS...91c4006C} using the \textsc{genec} code. The luminosity of the 3D model  was provided by a parameterised nuclear energy generation  rate, energy losses were also accounted for through escaping neutrinos, using a specific neutrino production rate, although their effect was negligible. The luminosity of the model was boosted by a factor of 1,000 \corr{in order to ease the time needed to establish the turbulent velocity field}, as discussed in \S\ref{comp-dom}. The computational domain utilised a plane-parallel geometry within a Cartesian coordinate system and used a parameterised gravitational acceleration. 

\par We tested the dependence of our set-up on the domain mesh size by computing models of four different resolutions: $128^3$, $256^3$, $512^3$ and $1024^3$. At these resolutions, approximate numerical Reynolds number of 650, 1600, 4000 and 10$^4$, respectively are achieved in the convective zone (using Eq. \ref{re_eff}). This means that with the exception of the \textsf{lrez} model, all of the models reach the turbulent regime (Re$_{eff}$ $\gtrsim 1000$). While a resolution of $512^3$ appears to produce a converged result at the upper boundary; the stiffer, lower boundary continues to change up to our highest resolution model. An even higher resolution run is thus planned. 

\par We observed entrainment of material at both convective boundaries for all of the models considered. This entrainment \corr{over the quasi-steady turbulent state is associated with an almost constant velocity, and the corresponding} time-scale is greater than the time-scale for the largest fluid elements to transit the convective region, asserting that convective 
boundary mixing in these models occurs within the equilibrium entrainment regime. The average 
entrainment velocities over the respective boundaries are $2.2\times10^4\,$cm$\,$s$^{-1}$ and 
$-0.3\times10^4\,$cm$\,$s$^{-1}$ for the upper and lower boundary, respectively. We also found that the entrainment 
velocity scales with the stiffness 
(bulk Richardson number) of the convective boundaries. This scaling follows the entrainment law with entrainment 
coefficient and exponent, \corr{$A=0.03\,(\pm0.01)$ and $n=0.62\,(+0.09/-0.15)$, respectively. 
These constants were obtained from only two convective boundaries. Additional simulations using different initial conditions should help explore the parameter space of the entrainment law and whether or not the parameters we derived vary significantly from one burning stage to the other. Furthermore, the dependence on the P\'eclet number needs to be further explored before our results obtained in the neutrino-cooled advanced phases can be applied to the early phases (hydrogen and helium burning) during which thermal effects are important, at least at the small scales \citep[see discussion in][]{2015A&A...580A..61V}.}
We also estimated the boundary widths and found these to be roughly 30\% and 10\% of the local pressure scale height for 
the upper and lower convective boundary, respectively. While these widths are only estimates, they confirm that the 
lower boundary is narrower than the upper boundary.

Although more 3D simulations of all burning stages are needed to fully \corr{characterize convective boundary mixing}, we can already compare our results to those of previous studies as well as the 1D input stellar model and relate them via measures of the 
turbulent driving and boundary stiffness. For 
this purpose, we investigated how entrainment and turbulent velocities, the driving luminosity and boundary stiffness 
(measured using the bulk Richardson number) relate to each other in \S\ref{comp1d3d}. Considering these 
relations enabled us to reconcile the convective boundary properties of the carbon shell estimated from the initial 1D stellar 
evolution model to the properties of boundaries in the 3D simulations presented here (despite the artificial increase 
in luminosity for the 3D simulations). Referring to the similarities between carbon and oxygen shell simulations presented in 
\S\ref{other_sims}, {\em a coherent picture seems to emerge from all existing simulations 
related to the advanced burning stages in massive stars when considering the relations between the above quantities. }

\par This is promising for the long term goal of developing a convective boundary mixing prescription for 1D models 
which is applicable to all (or many) stages of the evolution of stars (and not only to the specific conditions studied in 3D simulations). The 
luminosity (driving convection) and the bulk Richardson number (a measure of the boundary stiffness) will  
be key quantities for such new prescriptions \citep[also see][]{2015ApJ...809...30A}. 

The goal of 1D stellar evolution models is to capture the long-term (secular) evolution of the convective zone and of 
its boundaries, while 3D hydrodynamic simulations probe the short-term (dynamical) evolution. Keeping this in mind, the 
key points to take from this and previous 3D hydrodynamic studies for the development of new prescriptions in 1D 
stellar evolution codes are the following:
\vspace{-0.1cm}
\begin{itemize}
 \item Entrainment of the boundary and mixing across it occurs both at the top and bottom boundaries. Thus 1D 
stellar evolution models should include convective boundary mixing at both boundaries. Furthermore, the boundary shape 
is not a discontinuity in the 3D hydrodynamic simulations but a smooth function of radius, sigmoid-like, a feature that 
should also be incorporated in 1D models.
 \item At the lower boundary, which is stiffer, the entrainment is slower and the boundary width is narrower. This 
confirms the dependence of entrainment and mixing on the stiffness of the boundary. 
 \item Since the boundary stiffness varies both in time and with the convective boundary considered, a single constant 
parameter is probably not going to correctly represent the dependence of the mixing on the instantaneous convective 
boundary properties. As discussed above, we suggest the 
use of the bulk Richardson number in new prescriptions to include this dependence. 
\end{itemize}
\vspace{4cm}

\section*{Acknowledgements}
The authors acknowledge support from EU-FP7-ERC-2012-St Grant 306901. RH acknowledges support from the World Premier International Research Centre Initiative (WPI Initiative), MEXT, Japan. 
This article is based upon work from the “ChETEC” COST Action (CA16117), supported by COST (European Cooperation in Science and Technology). 
This work used the Extreme Science and Engineering Discovery Environment (XSEDE), which is supported by National Science Foundation grant number OCI-1053575. CM and WDA acknowledge support from NSF grant 1107445 at the 
University of Arizona. The authors acknowledge the Texas Advanced Computing Center (TACC) at The University of Texas at Austin (http://www.tacc.utexas.edu) for providing HPC resources that have contributed 
to the research results reported within this paper. MV acknowledges support from the European Research Council through grant ERC-AdG No. 341157-COCO2CASA. This work used the DiRAC Data Centric system at 
Durham University, operated by the Institute for Computational Cosmology on behalf of the STFC DiRAC HPC Facility (www.dirac.ac.uk). This equipment was funded by BIS National E-infrastructure capital grant 
ST/K00042X/1, STFC capital grants ST/H008519/1 and ST/K00087X/1, STFC DiRAC Operations grant ST/K003267/1 and Durham University. DiRAC is part of the National E-Infrastructure.

\newpage

\begin{appendix}
\section{1D Stellar Convection Parameters}\label{convection-parameters-appendix}

\begin{table*}
\begin{center}
\caption{Estimates of the convective velocity (cm$\,$s$^{-1}$), Bulk Richardson number, Mach number, P\'{e}clet 
number and Damk\"{o}hler number of different times during core and shell burning phases of a 15$\,\textrm{M}_\odot$ 
stellar model. Bulk Richardson numbers are boundary values, brackets indicate values at the lower boundary, all other 
values were mass averaged over the convective region. P\'{e}clet numbers are order of magnitude 
estimates.}\begin{tabulary}{\textwidth}{l || c c c c c}
\hline \hline\\
\textsf{Phase} & \textsf{$v_c$} (cm$\,$s$^{-1}$) & \textsf{Ri$_{\rm B}$} & \textsf{Ma} & \textsf{Pe} & \textsf{Da} \\\\
\hline\hline\\
\textsf{ H Core Start } &  6.9$\times10^{4}$  &  1.8$\times10^{2}$  &  9.3$\times10^{-4}$  &  $\sim10^{3}$  &  
3.8$\times10^{-8}$ \\
\textsf{ H Core End } &  9.7$\times10^{4}$  &  1.1$\times10^{2}$  &  1.5$\times10^{-3}$  &  $\sim10^{3}$  &  
7.2$\times10^{-7}$ \\\\
\hline\\
\textsf{ He Core Start } &  4.7$\times10^{4}$  &  1.2$\times10^{3}$  &  4.3$\times10^{-4}$  &  $\sim10^{4}$  &  
1.8$\times10^{-7}$ \\
\textsf{ He Core Max } &  5.9$\times10^{4}$  &  4.0$\times10^{2}$  &  5.0$\times10^{-4}$  &  $\sim10^{5}$  &  
3.5$\times10^{-6}$ \\
\textsf{ He Core End } &  5.9$\times10^{4}$  &  3.8$\times10^{2}$  &  5.1$\times10^{-4}$  &  $\sim10^{5}$  &  
3.0$\times10^{-6}$ \\\\
\hline\\
\textsf{ C Core Start } &  6.9$\times10^{4}$  &  7.2$\times10^{3}$  &  3.0$\times10^{-4}$  &  $\sim10^{6}$  &  
3.4$\times10^{-6}$ \\
\textsf{ C Core Max } &  5.6$\times10^{4}$  &  1.2$\times10^{4}$  &  2.4$\times10^{-4}$  &  $\sim10^{7}$  &  1.2$\times10^{-5}$ \\
\textsf{ C Core End } &  5.8$\times10^{4}$  &  3.9$\times10^{2}$  &  2.4$\times10^{-4}$  &  $\sim10^{7}$  &  
1.8$\times10^{-5}$ \\\\
\hline\\
\textsf{ Ne Core Start } &  1.5$\times10^{6}$  &  82  &  3.9$\times10^{-3}$  &  $\sim10^{10}$  &  3.3$\times10^{-3}$ \\
\textsf{ Ne Core Max } &  6.4$\times10^{5}$  &  3.6$\times10^{2}$  &  1.9$\times10^{-3}$  &  $\sim10^{10}$  &  
5.5$\times10^{-3}$ \\
\textsf{ Ne Core End } &  4.1$\times10^{5}$  &  62  &  1.1$\times10^{-3}$  &  $\sim10^{10}$  &  2.6$\times10^{-3}$ \\\\
\hline\\
\textsf{ O Core Start } &  8.8$\times10^{5}$  &  2.4$\times10^{2}$  &  2.2$\times10^{-3}$  &  $\sim10^{10}$  &  
6.3$\times10^{-4}$ \\
\textsf{ O Core Max } &  7.9$\times10^{5}$  &  8.5$\times10^{4}$  &  2.0$\times10^{-3}$  &  $\sim10^{10}$  &  2.4$\times10^{-3}$ \\
\textsf{ O Core End } &  7.5$\times10^{5}$  &  27  &  1.8$\times10^{-3}$  &  $\sim10^{10}$  &  2.0$\times10^{-3}$ \\\\
\hline\\
\textsf{ He Shell Start } &  1.4$\times10^{5}$  &  46  ( 20 ) &  8.9$\times10^{-4}$  &  $\sim10^{6}$  &  5.7$\times10^{-8}$ \\
\textsf{ He Shell End } &  1.3$\times10^{5}$  &  14  ( 1.8$\times10^{3}$ ) &  9.1$\times10^{-4}$  &  $\sim10^{6}$  &  
1.0$\times10^{-7}$ \\\\
\hline\\
\textsf{ C Shell Start } &  3.6$\times10^{5}$  &  4.2$\times10^{2}$  ( 6.0$\times10^{3}$ ) &  1.3$\times10^{-3}$  &  
$\sim10^{8}$  &  1.3$\times10^{-4}$ \\
\textsf{ C Shell IC$^a$ } &  2.9$\times10^{5}$  &  6.9$\times10^{2}$  ( 1.5$\times10^{4}$ ) &  
1.2$\times10^{-3}$  &  $\sim10^{8}$  &  2.0$\times10^{-4}$ \\
\textsf{ C Shell End } &  1.6$\times10^{5}$  &  59  ( 6.5$\times10^{4}$ ) &  5.7$\times10^{-4}$  &  
$\sim10^{7}$  &  1.3$\times10^{-4}$ \\\\
\hline\\
\textsf{ O Shell Start } &  1.5$\times10^{5}$  &  3.7$\times10^{4}$  ( 4.0$\times10^{4}$ ) &  3.4$\times10^{-4}$  &  $\sim10^{10}$  &  2.7$\times10^{-4}$ \\
\textsf{ O Shell End } &  5.7$\times10^{5}$  &  1.2$\times10^{2}$  ( 3.4$\times10^{4}$ ) &  1.3$\times10^{-3}$  &  
$\sim10^{10}$  &  1.4$\times10^{-3}$
\label{param}
\end{tabulary}
\end{center}
\ \\ \ \\ $^a$ Properties of the 1D model used as initial conditions for the 3D simulations
\end{table*}

\par To determine averages over the convective region, we used an RMS mass average. The mass average of a quantity $A$ is defined as:

\begin{equation}\label{mavg}
A_{\rm avg}=\sqrt{\frac{1}{m_2-m_1}\int_{m_1}^{m_2}A^2(m)\, dm},
\end{equation}

\noindent where $m_1$ and $m_2$ are the mass at the lower and upper convective boundaries, respectively.

The bulk Richardson number is the ratio of the boundary stabilising potential to the kinetic energy of turbulent motions 
(within the convective region). It characterises the boundary stiffness, and is a function of the Brunt-V\"{a}is\"{a}l\"{a} frequency or buoyancy frequency, $N$, which is defined as:

\begin{equation}\label{n2}
N^2 = -g\left(\frac{\partial \textrm{ln} \rho}{\partial r}\Big|_e - \frac{\partial \textrm{ln} \rho}{\partial r}\Big|_s\right),
\end{equation}

\noindent where $g$ is the gravitational acceleration, $\rho$ is the density, and subscripts $e$ and $s$ represent the fluid element and its surroundings, respectively. 

In order to study the properties of the boundary regions, we first need a definition of the boundary location, $r_c$. 
\citet{2007ApJ...667..448M} use the maximum gradient method on the 
composition to determine the location of convective boundaries in the oxygen shell of a
23$\,\textrm{M}_\odot$ model. In a similar manner, we approximate the location of the top (bottom) boundary
as the vertical coordinate having an average atomic mass, $\overline{A}$, equal to the average between the convective
zone and top (bottom) radiative zone,

\begin{equation}\label{boundaryloc}
\overline{A}_{th}=\frac{\overline{A}_{\rm conv}+\overline{A}_{\rm rad}}{2},
\end{equation}

\noindent where $\overline{A}_{conv}$ and $\overline{A}_{rad}$ are the averages of $\overline{A}$ in the convective
and relevant stable/radiative regions, respectively. 

The buoyancy jump over a convective boundary region can be estimated by integrating the 
square of the buoyancy frequency over a suitable distance ($\Delta r$) either side of the boundary centre, 
$r_c$,

\begin{equation}\label{buoy}
\Delta B=\int\limits_{r_c-\Delta r}^{r_c+\Delta r}N^2dr.
\end{equation}

The integration distance $\Delta r$ is not well defined theoretically but it should be large enough to capture the dynamics of the boundary region and the distance over which fluid 
elements are decelerated. 

As mentioned above, the bulk Richardson number is the ratio of the boundary stabilisation potential (which includes the buoyancy jump) to the kinetic energy due to turbulent motions 
(within the convective region),

\begin{equation}\label{rib}
\textrm{Ri}_{\rm B} = \frac{\Delta B \ell}{v_{rms}\,^2},
\end{equation}

\noindent where $\ell$ is the integral length-scale which represents the size of the largest fluid elements. The integral length-scale is often taken to be the horizontal correlation length. 
\citet{2007ApJ...667..448M} show that the horizontal correlation length-scale and pressure scale height are similar to within a factor $\sim 3$. So for our analysis we use the pressure scale height close 
to the boundary. The RMS velocity, $v_{rms}$, represents the velocity of the largest fluid elements carrying most of the 
energy, which for the 1D simulations we approximate as the convective velocity, 

\begin{equation}\label{vc}
v_c=\left(\frac{F_c}{\rho}\right)^{\frac{1}{3}},
\end{equation}

\noindent where $F_c$ is the convective flux.

\par In estimating the Mach number, we determine the sound speed, $c_s$, using the Helmholtz EOS
\citep{1999ApJS..125..277T,2000ApJS..126..501T},

\begin{equation}\label{ma}
\textrm{Ma} = \frac{v_c}{c_s}.
\end{equation}

The P\'{e}clet number (Pe) is defined as the ratio of the time-scale for advective transport to the time-scale for 
transport through diffusion. In the stellar case thermal diffusion dominates over molecular diffusion. For the deep 
interior heat transfer plays a minor role, so typically Pe $>>1$. We determine the P\'{e}clet number using the 
following formula,

\begin{equation}\label{pe}
\textrm{Pe} = \frac{3 D_{mlt}}{\chi},
\end{equation}

\begin{table*}
\captionof{table}{Constants used in the fitting functions (Eq. \ref{fit_fun}) for the five sections of the entropy, 
$\bar{A}$ and $\bar{Z}$ profiles. Subscripts $1,2$ and $3$ refer to the lower stable, convective and upper 
stable sections, respectively. Subscripts $l$ and $u$ refer to the lower and upper convective 
boundary sections, respectively.}
\begin{tabulary}{\textwidth}{l || c c c c c c c c c c c c}
\hline \hline\\
 & \textsf{$\alpha_1$} & \textsf{$\beta_1$} & \textsf{$\theta_l$} & \textsf{$\phi_l$} & 
\textsf{$\eta_l$} & \textsf{$\alpha_2$} & \textsf{$\beta_2$} & \textsf{$\theta_u$} & 
\textsf{$\phi_u$} & 
\textsf{$\eta_u$} & \textsf{$\alpha_3$} & \textsf{$\beta_3$}\\\\
\hline \hline\\
\textsf{$s$}  &  $1.65\times10^8$  &  $0.24$  &  $2.81\times10^8$  &  $3.43\times10^8$  
&  $-0.5$  &  $3.43\times10^8$  &  $0$  &  $3.43\times10^8$  &  $3.56\times10^8$  &  $-0.5$  &  
$3.56\times10^8$  &  $0.08$  \\\\
\textsf{$\bar{A}$}  &  $18.17$  &  $0$  &  $18.17$  &  $16.19$  &  $0.5$  &  $16.19$  & 
 0  &  $16.19$  &  $14.38$  &  $0.5$  &  $14.38$  &  $0$ \\\\
\textsf{$\bar{Z}$}  &  $9.07$  &  $0$  &  $9.07$  &  $8.08$  &  $0.5$  &  $8.08$  &  0  
&  $8.08$  &  $7.18$  &  $0.5$  &  $7.18$  &  $0$ \\\\
\hline \hline
\label{const1d}
\end{tabulary}
\end{table*}

\noindent where $D_{mlt}$ is the diffusion coefficient calculated using MLT, $D_{mlt}=v_{mlt}\ell_{mlt}/3$, $v_{mlt}$ and 
$\ell_{mlt}$ are the mixing length velocity and mixing length parameter. $\chi$ is the thermal diffusivity, defined as,

\begin{equation}\label{chi}
\chi=\frac{16 \sigma\, T^3}{3\kappa\,\rho^2\,c_p}
\end{equation}

\noindent where $\sigma$ is the Stefan-Boltzmann constant, $T$ the temperature, $\kappa$ the Rosseland mean opacity and $c_p$ the specific heat capacity at constant pressure.

The Damk\"{o}hler number (Da) is defined as the ratio of the advective time-scale to the nuclear reaction time-scale, 
generally this is small in deep convective regions as the time-scale for nuclear reactions is long, but during the 
advanced stages of massive star evolution the two time-scales can become comparable. We determine the Damk\"{o}hler 
number using the following formula,

\begin{equation}\label{da}
\textrm{Da} = \frac{t_{\rm con}}{t_{\rm nuc}} = \left(\frac{2 \ell_{c}}{v_c}\right)\left(\frac{q 
X_i}{\epsilon_{nuc}}\right)^{-1},
\end{equation}

\noindent where $t_{\rm con}$ is the convective turnover time and $t_{\rm nuc}$ is the relevant nuclear reaction time-scale. 
$\ell_{c}$ is the height of the convective zone, $q$ is the specific energy released for the dominating reactions, 
$X_i$ is the mass abundance of the interacting particles and $\epsilon_{nuc}$ is the nuclear energy generation rate.

The majority of these variables are presented in Table \ref{param} for most of the convective boundaries shown in Fig. \ref{kipp}.

\section{Stellar Model Profile Fitting}\label{app}

\noindent The entropy ($s$), average atomic mass ($\bar{A}$) and average atomic number ($\bar{Z}$) were 
remapped by considering five distinct sections of the domain. The lower stable region (below the lower 
convective boundary), the convective region and the upper stable region (above the upper boundary) 
were fitted linearly in the form $\alpha+\beta\,x$, where $\alpha$ and $\beta$ are constants, and $x$ is 
the radius on a grid point. The two remaining sections are the upper and lower convective 
boundaries, these were fitted using sigmoid functions, $f_{sig}$, of the form,

\begin{equation}\label{fit_fun}
f_{sig}=\theta+\frac{\phi-\theta}{1+e^{\,\eta\,z}}, 
\end{equation}

where $\theta$, $\phi$ and $\eta$ are constants, and $z$ is a normalised grid index. The fitting 
constants for the three variables are presented in Table \ref{const1d}, the subscripts for each constant 
represent the section of the domain for which the fit refers to. Subscripts $1,2,3$ denote the lower 
stable, convective and upper stable sections, respectively. Subscripts $l$ and $u$ refer to the lower 
and upper convective boundary sections, respectively.


\section{RANS terminology and the Turbulent Kinetic Energy Equation}\label{app-tke}

In the Reynolds-averaged Navier Stokes (RANS) framework, variables are split into mean and 
fluctuating components. The
horizontally averaged mean is denoted by angled brackets and defined as,

\begin{equation}
\left<a\right> = \frac{1}{\Delta A}\int_{\Delta A}a \,dA,
\end{equation}
where $\,dA=dy\,dz\,$ and $\,\Delta A = \Delta y \Delta z\,$ is the area of the computational domain. 
The fluctuating component, $a'$, is obtained by subtracting the mean of the variable from the variable:  $a'= a - \left<a\right>$.

In order to statistically sample the quasi-steady state we perform temporal averaging over several convective turnovers, denoted by an over-bar and defined as,

\begin{equation}
\overline{a}=\frac{1}{\Delta t}\int_{t_1}^{t_2}a(t)\, dt,
\end{equation}
\noindent for an averaging window $\Delta t=t_2-t_1$. 

\par The Eulerian equation of turbulent kinetic energy can be written as (eq. A12 of \citealt{2007ApJ...667..448M}):
\begin{equation} \label{mfka}
\partial_t\left(\rho E_k\right)+ \boldsymbol{\nabla}\cdot\left(\rho E_k\boldsymbol{v}\right)=-\boldsymbol{v}\cdot\boldsymbol{\nabla}p+\rho\,\boldsymbol{v}\cdot\boldsymbol{g}\\
\end{equation}

\noindent where $\boldsymbol{v}$ is the velocity and $E_k=\frac{1}{2}(\boldsymbol{v}\cdot\boldsymbol{v})$ is the specific kinetic energy.\\

Applying horizontal and temporal averaging to Eq. \ref{mfka} yields the mean turbulent kinetic energy equation,
which can be written as,

\begin{align}
\begin{split} \label{mmfka}
\overline{\left<\rho \mathbf{D_t} E_k\right>}=&-\boldsymbol{\nabla}\cdot\overline{\left<\mathbf{F_p}+\mathbf{F_k}\right>}\\
&+\overline{\left<\mathbf{W_p}\right>}+\overline{\left<\mathbf{W_b}\right>}-\epsilon_k,
\end{split}
\end{align}

where $\mathbf{D_t} =\partial_t + \boldsymbol{\nabla}\cdot\boldsymbol{v}$ is the material derivative,

$\mathbf{F_p}=p'\boldsymbol{v'}$ is the turbulent pressure flux,

$\mathbf{F_k}=\rho E_k\boldsymbol{v'}$ is the turbulent kinetic energy flux,

$\mathbf{W_p}=p'\boldsymbol{\nabla}\cdot\boldsymbol{v'}$ is the pressure dilatation,

$\mathbf{W_b}=\rho'\boldsymbol{g}\cdot\boldsymbol{v'}$ is the work due to buoyancy and

$\epsilon_k$ is the numerical dissipation of kinetic energy.\\

\end{appendix}

\clearpage
\bibliography{references}

\bsp

\label{lastpage}
\end{document}